\documentclass[journal, twocolumn]{IEEEtran}

\usepackage[T1]{fontenc}
\usepackage{courier}

\usepackage{amsmath,amssymb,amsfonts}
\newcommand{\ignore}[1]{ }
\usepackage{graphicx}
\usepackage{textcomp}
\usepackage[table]{xcolor}
\usepackage{float}
\usepackage{booktabs}
\usepackage{multirow}
\usepackage{array}
\usepackage{adjustbox}
\usepackage{longtable}
\usepackage{lipsum}
\usepackage{pifont}
\usepackage{cite}
\usepackage{soul}
\usepackage{ragged2e}
\usepackage{setspace}
\usepackage{tikz}

\usepackage[colorlinks=true, linkcolor=blue, citecolor=blue, urlcolor=blue, filecolor=blue]{hyperref}
\definecolor{headerblue}{RGB}{31,73,125}  
\usepackage[disable]{todonotes}

\newcommand{\red}{}

\renewcommand{\red}{\textcolor{red}}

\newcommand{\QoE}{\textbf{\texttt{QoE}}}

\newcounter{mysfig}
\renewcommand\themysfig{\thefigure(\alph{mysfig})}
\makeatletter
\newcommand\Scaption[1]{%
\vskip.5\abovecaptionskip
  \sbox\@tempboxa{\small~#1}%
  \ifdim \wd\@tempboxa >\hsize
    \small\themysfig~#1\par
  \else
    \global \@minipagefalse
    \hb@xt@\hsize{\hfil\box\@tempboxa\hfil}%
  \fi
  \vskip\belowcaptionskip}
\makeatother

\definecolor{lime}{HTML}{A6CE39}
\DeclareRobustCommand{\orcidicon}{%
	\begin{tikzpicture}
	\draw[lime, fill=lime] (0,0)
	circle [radius=0.16]
	node[white] {{\fontfamily{qag}\selectfont \tiny ID}};
	\draw[white, fill=white] (-0.0625,0.095)
	circle [radius=0.007];
	\end{tikzpicture}
	\hspace{-2mm}
}

\foreach \x in {A, ..., Z}{%
	\expandafter\xdef\csname orcid\x\endcsname{\noexpand\href{https://orcid.org/\csname orcidauthor\x\endcsname}{\noexpand\orcidicon}}
}

\begin{document}


\title{Enhancing Quality of Experience in Future Telecommunication Networks:
A Review of Frameworks and Machine Learning Approaches}

\author{
\IEEEauthorblockN{Parsa Hassani Shariat Panahi\IEEEauthorrefmark{1},
                  Amir H. Jalilvand\IEEEauthorrefmark{1},
                  M. Hassan Najafi\IEEEauthorrefmark{2}\\}
\IEEEauthorblockA{\IEEEauthorrefmark{1}School of Computer Engineering,
                  Iran University of Science and Technology, Tehran, Iran}
\IEEEauthorblockA{\IEEEauthorrefmark{2}Department of Electrical, Computer and
                  Systems Engineering,
                  Case Western Reserve University, Cleveland, OH, USA}
}

\maketitle

\begin{abstract}
Quality of experience (\QoE{}) has emerged as a key measure of user satisfaction
for multimedia services, including adaptive video streaming, immersive media
(AR/VR) and cloud gaming. 
As services grow more varied, \QoE{} \textit{assessment}
has to 
connect \textit{measurement} with the \textit{models} that act on it. Existing surveys treat
the two as separate catalogues and do not classify how they are \textit{coupled},
so the choice of an integrated architecture is neither replicable nor
transferable. This survey addresses four questions: (1) how frameworks are coupled
with ML models, (2) which \QoE{} actions each pattern supports,  (3) which
patterns appear in each service domain, and (4) which problems remain open.  A taxonomy
of loose, tight, and distributed/federated coupling answers them, in which each
work is coded by where model inference runs. This paper also identifies
three enhancement mechanisms (predictive optimization, adaptive resource
allocation, content-aware enhancement). Because reported accuracies use different metrics and datasets, they cannot be
ranked directly. A model-selection guide is therefore derived from verified
results. The study traces three generations,
from static measurement through reactive ML to proactive and generative methods,
including large language model integration, and examines domain constraints for
immersive media, cloud gaming and the Internet of Things. Open challenges include
generalizability, interpretability, model drift and privacy, while objective
tuning, federated scheduling, multi-modal and mobile-edge integration are
emerging directions.
\end{abstract}
\begin{IEEEkeywords}
Quality of experience (\textcolor{black}{\QoE{}}), multimedia services, machine learning (ML),
measurement frameworks, integration patterns, immersive media, cloud gaming,
Internet of Things (IoT), explainable AI.
\end{IEEEkeywords}

\vspace{-1em}
\section{Introduction}
\label{sec:intro}

\begin{figure*}
    \centering
    \includegraphics[width=\linewidth, trim=0.2cm 0cm 0cm 0cm, clip]{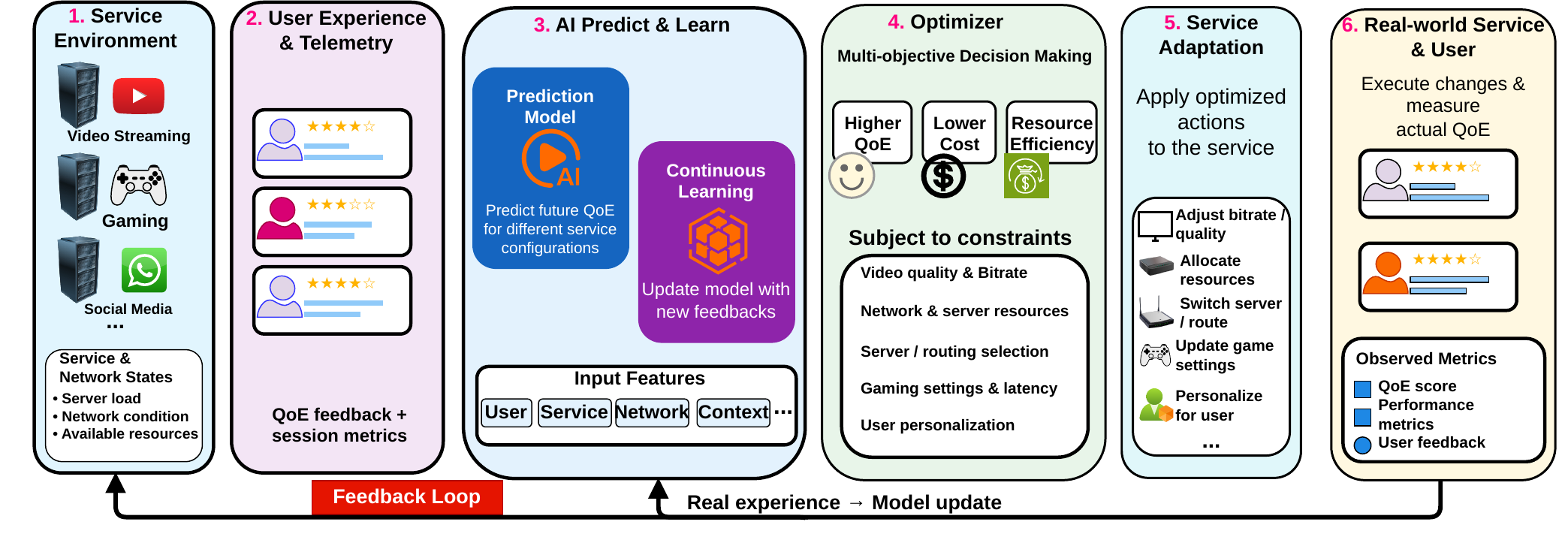}
    \vspace{-1.5em}
    \caption{Reference closed loop for \QoE{} prediction and optimization, used as the organizing frame of this survey. The system integrates service monitoring, user experience and telemetry collection, AI-based \QoE{} prediction, multi-objective optimization, and service adaptation. The optimizer selects feasible service configurations that balance predicted \QoE{} improvement against operational costs and resource constraints. Real-world \QoE{} measurements and performance telemetry feed back into the learning component to support continuous model adaptation, closing the loop between prediction, decision-making, service execution and observed user experience. Stages 2 and 3 are reviewed in \autoref{sec:landscape}; the link between them (\texttt{RQ1}) in \autoref{sec:patterns}; stages 4 and 5 (\texttt{RQ2}) in \autoref{sec:mechanisms}; stage 1 (\texttt{RQ3}) in \autoref{sec:domains}; stage 6 (\texttt{RQ4}) in \autoref{sec:challenge}.}
    \label{fig:all-in-one}
    \vspace{-1.5em}
\end{figure*}

\IEEEPARstart{M}{obile} data traffic, projected to exceed 600 exabytes per month by
2030~\cite{j1-2}, drives heterogeneous applications including video streaming,
AR/VR, cloud gaming and the Internet of Things (IoT), each with distinct
latency, bandwidth and reliability requirements~\cite{new-1,new-2}. Traditional
quality of service (QoS) frameworks guarantee network-layer metrics but cannot
capture whether users are satisfied with the services they
receive~\cite{4624234}. Quality of experience (\QoE{}) addresses this limitation by
modeling user satisfaction as a function of system, user and context
factors~\cite{new-4,new-5,kougioumtzidis2022survey}. Accurate \QoE{} assessment has
become a priority for network operators, content providers and standardization
bodies~\cite{new-9,new-10,p1-3}. Both closed-source commercial platforms and
open-source research tools have been developed to collect objective and
subjective \QoE{} data. In parallel, machine learning (ML) has become a common
approach to \QoE{} prediction, modeling non-linear relationships between network
KPIs and perceptual quality scores that statistical methods cannot
represent~\cite{ml-9,ml-6}. The combination of scalable measurement
infrastructure and data-driven ML is the current focus of \QoE{} management for
multimedia services, as illustrated in \autoref{fig:all-in-one}.

Despite this progress, existing efforts follow three separate lines of work.
Media-quality-assessment work, from large-model quality-assessment
surveys~\cite{zhang2025quality} to large language model (LLM)-based video-quality
methods~\cite{ge2024lmmvqa}, scores perceptual quality rather than the systems
that produce and act on those scores. ML-based \QoE{}-modeling and
adaptive-streaming
surveys~\cite{ghosh2024review,kougioumtzidis2022survey,nasralla2023exploring}
catalog which models predict \QoE{} and how to adapt streaming, while
immersive-media studies characterize perceptual quality and visual attention for
emerging formats such as dynamic point clouds~\cite{zhou2025pointcloud}. The link
between open- or closed-source \QoE{} measurement frameworks and the ML models
that act on their output is not addressed.
Because each of these lines treats measurement and modeling separately, four
gaps remain. To our knowledge, no survey classifies the architectural coupling
patterns between measurement frameworks and ML, so practitioners have no guidance
on latency, scalability, or privacy trade-offs~\cite{kougioumtzidis2022survey}.

In this paper, \QoE{} measurement frameworks and ML models are treated as one
integrated system rather than as two separate catalogues, and the reference loop
in \autoref{fig:all-in-one} frames the analysis. Four research questions follow
from this view: \texttt{RQ1} asks how \QoE{} measurement frameworks are coupled with ML
models; \texttt{RQ2} asks which \QoE{} actions each coupling pattern supports; \texttt{RQ3} asks
which coupling patterns have been applied in each service domain and which
service constraints explain these choices; and \texttt{RQ4} asks which problems remain
open. \texttt{RQ1} concerns the link between stages 2 and 3 of the loop, \texttt{RQ2} stages 4
and 5, \texttt{RQ3} the service environment of stage 1, and \texttt{RQ4} the feedback of stage 6.

The contributions of this survey answer these four questions. First, a taxonomy
of framework-ML coupling with three patterns: \emph{loose coupling}, \emph{tight
coupling} and \emph{distributed/federated} integration. The pattern is decided by
where model inference runs relative to the measurement loop. Most ML-based works
train and test offline; only two report evidence from a production service (\texttt{RQ1},
\autoref{sec:patterns}). Second, a mapping of the \QoE{} actions in the reviewed
works to the coupling patterns that enable them, grouped into \emph{predictive
optimization}, \emph{adaptive resource allocation} and \emph{content-aware
enhancement}, with the trade-offs between real-time response, scalability and
privacy for each pattern (\texttt{RQ2}, \autoref{sec:mechanisms}). Third, a domain
analysis of adaptive video streaming, live streaming, immersive media (AR/VR),
cloud gaming and IoT, which links each domain's coupling to its service
constraints and gives a model-selection guide by latency, dataset size and model
complexity (\texttt{RQ3}, \autoref{sec:domains}). Fourth, a set of open challenges,
including model drift, interpretability, privacy and benchmarking, together with
five emerging directions that rest on a single work each: objective tuning,
near-real-time RAN intelligent controller (RIC)-mediated control, federated
scheduling, multi-modal integration and mobile-edge integration (\texttt{RQ4},
\autoref{sec:challenge}).

The remainder of this paper follows the loop of \autoref{fig:all-in-one}.
\autoref{sec:background} introduces the \QoE{} fundamentals and the notation
used throughout. \autoref{sec:related} positions this work against prior surveys
and states the gap. \autoref{sec:method} describes how the reviewed works were
found, selected and coded. \autoref{sec:landscape} reviews the measurement
frameworks and ML models of stages 2 and 3. \autoref{sec:patterns} answers \texttt{RQ1}.
\autoref{sec:mechanisms} answers \texttt{RQ2}. \autoref{sec:domains} answers \texttt{RQ3}.
\autoref{sec:challenge} answers \texttt{RQ4}. \autoref{sec:conc} answers the four
questions in order. Abbreviations are listed in \autoref{tab:abbreviations}.
\begin{table*}[t]
\centering
\caption{{List of abbreviations used throughout this survey, grouped by
category. Terms are listed in their order of first appearance in the paper.
The dagger marks an expansion that has to be verified against the source before
submission.}}
\label{tab:abbreviations}
\renewcommand{\arraystretch}{1.1}
\setlength{\tabcolsep}{4pt}
\footnotesize
\rowcolors{2}{gray!10}{white}
\begin{tabular}{@{}>{\raggedright\arraybackslash}m{0.8cm}
                  >{\raggedright\arraybackslash}m{3.1cm}
                  >{\raggedright\arraybackslash}m{0.8cm}
                  >{\raggedright\arraybackslash}m{3.1cm}
                  >{\raggedright\arraybackslash}m{0.8cm}
                  >{\raggedright\arraybackslash}m{3.1cm}
                  >{\raggedright\arraybackslash}m{0.8cm}
                  >{\raggedright\arraybackslash}m{3.1cm}@{}}
\toprule
\rowcolor{headerblue}
{\color{white}\textbf{Term}} & {\color{white}\textbf{Meaning}} &
{\color{white}\textbf{Term}} & {\color{white}\textbf{Meaning}} &
{\color{white}\textbf{Term}} & {\color{white}\textbf{Meaning}} &
{\color{white}\textbf{Term}} & {\color{white}\textbf{Meaning}} \\
\midrule
QoS  & Quality of Service                & \QoE{}  & Quality of Experience             & VoIP & Voice over IP                      & AR   & Augmented Reality \\
VR   & Virtual Reality                   & MNO  & Mobile Network Operator           & LTE  & Long Term Evolution                 & MOS  & Mean Opinion Score \\
HTTP & Hypertext Transfer Protocol       & IoT  & Internet of Things                & ITU  & International Telecommunication Union & KPI & Key Performance Indicator \\
BS   & Base Station                      & UE   & User Equipment                    & CDN  & Content Delivery Network            & MEC  & Mobile Edge Computing \\
RAN  & Radio Access Network              & CFA  & Critical Feature Analytics$^{\dagger}$ & ML   & Machine Learning                    & SVR  & Support Vector Regression \\
SVM  & Support Vector Machine            & DT   & Decision Tree                     & RFR  & Random Forest Regression            & RNN  & Recurrent Neural Network \\
LSTM & Long Short-Term Memory            & CNN  & Convolutional Neural Network      & 3D CNN & 3D Convolutional Neural Network   & DNN  & Deep Neural Network \\
RBFN & Radial Basis Function Network     & KNN  & K-Nearest Neighbor                & LR   & Linear Regression                   & RR   & Ridge Regression \\
GPR  & Gaussian Process Regression       & ANN  & Artificial Neural Network         & DL   & Deep Learning                       & TCN  & Temporal Convolutional Network \\
GAN  & Generative Adversarial Network    & AI   & Artificial Intelligence           & 4G   & Fourth-Generation Network           & 5G   & Fifth-Generation Network \\
6G   & Sixth-Generation Network          & ABR  & Adaptive Bitrate                  & DASH & Dynamic Adaptive Streaming over HTTP & HAS  & HTTP Adaptive Streaming \\
HAR  & HTTP Archive                      & IPTV & Internet Protocol Television      & HTML & Hypertext Markup Language           & SDN  & Software-Defined Networking \\
NFV  & Network Functions Virtualization  & SLA  & Service Level Agreement           & ACR  & Absolute Category Rating            & RL   & Reinforcement Learning \\
CL   & Continual Learning                & BT   & Boosted Trees                     & GA   & Genetic Algorithm                   & RF   & Random Forest \\
LLM  & Large Language Model              & VQA  & Video Quality Assessment          & AIGC & AI-Generated Content                & VMAF & Video Multimethod Assessment Fusion \\
PSNR & Peak Signal-to-Noise Ratio        & PCC  & Pearson Correlation Coefficient   & SCC  & Spearman Correlation Coefficient    & RMSE & Root Mean Square Error \\
RSSI & Received Signal Strength Indicator & RSRP & Reference Signal Received Power  & RSRQ & Reference Signal Received Quality   & CINR & Carrier-to-Interference-plus-Noise Ratio \\
XR   & Extended Reality                  & 6DoF & Six Degrees of Freedom            & GDPR & General Data Protection Regulation   & CCPA & California Consumer Privacy Act \\
XAI  & Explainable Artificial Intelligence & AR/VR & Augmented and Virtual Reality   & RIC  & RAN Intelligent Controller          & O-RAN & Open Radio Access Network \\
SMO  & Service Management and Orchestration & SAC & Soft Actor-Critic               & PPO  & Proximal Policy Optimization        & FHE  & Fully Homomorphic Encryption \\
NOMA & Non-Orthogonal Multiple Access    & SVC  & Scalable Video Coding             & HDT  & Human Digital Twin                  & UAV  & Unmanned Aerial Vehicle \\
CoT  & Chain-of-Thought                  & TTFA & Time-to-First-Answer              & SHAP & SHapley Additive exPlanations       & VVC  & Versatile Video Coding \\
\bottomrule
\end{tabular}
\vspace{-1em}
\end{table*}

\section{Background}
\label{sec:background}

Formally, \QoE{} is defined as a multidimensional function of system ($S$), user
($U$), and context ($C$) influence factors~\cite{kougioumtzidis2022survey}:
\begin{equation}
    \QoE{} = f(S, U, C).
    \label{eq:qoe}
\end{equation}
In the context of \QoE{}-driven resource management, a primary objective is to
maximize this utility function subject to resource
constraints~\cite{nasralla2023exploring}:
\begin{equation}
    \max_{x} \QoE{}(x) \quad \text{s.t.} \quad \mathrm{Cost}(x) \leq B,
    \label{eq:opt}
\end{equation}
where $x$ represents the vector of allocatable resources (\textit{e.g.,} bandwidth,
compute power) and $B$ is the available budget or capacity. While this
formulation focuses on cost-constrained optimization, practical \QoE{} management
has to balance multiple competing objectives, including capacity utilization, SLA
compliance, fairness across users, and energy efficiency. The cost-only
formulation is adopted here as a tractable baseline that captures the dominant
trade-off in \QoE{}-driven resource allocation~\cite{j1-27}.

\QoE{} impact factors include four categories. \emph{Human-related} factors
(demographics, emotional state, attention) are studied mostly through
demographics, with transient aspects
neglected~\cite{j1-5}. \emph{System-related} factors (delay, throughput, packet
loss, encoding) are extensively studied but often overlook interactions with user
perception~\cite{j1-27}. \emph{Context-related} factors (location, mobility,
social factors) add complexity, and current approaches do not account for dynamic
environments~\cite{j1-28}. \emph{Content-related} factors (encoding rate,
resolution, video quality) are often analyzed separately rather than as part of
an integrated framework~\cite{j1-26}.

\begin{figure*}
    \centering
    \includegraphics[width=0.9\linewidth,clip,trim={0cm 1cm 0cm 0cm}]{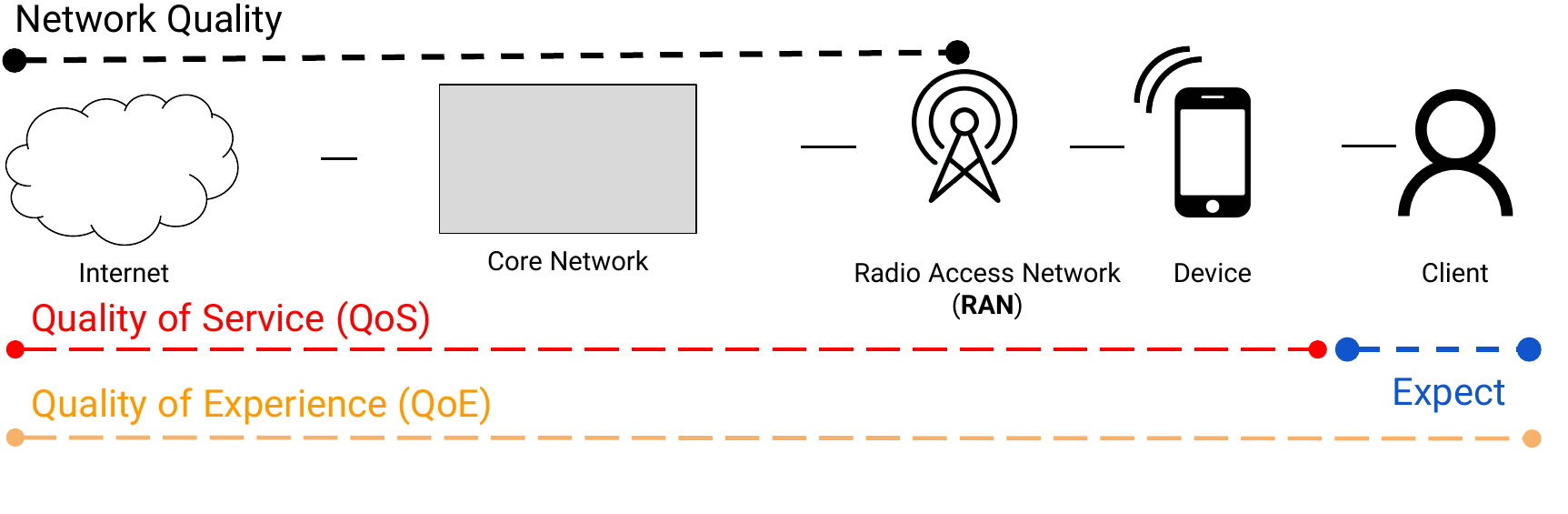}
    \vspace{-1em}
    \caption{Network quality, QoS, and \QoE{} domains within a cellular network
    infrastructure. QoS spans network layers 1--4, while \QoE{} spans layers 1--7.}
    \label{fig:qoe-vs-qos}
\vspace{-1em}
\end{figure*}

As shown in \autoref{fig:qoe-vs-qos}, QoS measures network-layer performance
(delay, jitter, packet loss, throughput) at layers 1--4, while \QoE{} spans layers
1--7 and incorporates user perception, expectations, and context. Good QoS does
not guarantee good \QoE{} (for example, a high bitrate with poor content), and poor
QoS can still yield acceptable \QoE{} (for example, through adaptive streaming).
These differences motivate cross-layer optimization~\cite{dif-1,dif-2}.

A parallel line of work standardizes how \QoE{} is measured. Recommendation
ITU-T G.1051~\cite{itut2023g1051} specifies a methodology for measuring the
\QoE{} of real-time interactive services in 5G and beyond-5G systems. The
methodology defines traffic patterns that emulate the client-server data flows
of a target service, measures a common set of QoS key performance indicators
(KPIs) such as round-trip time, packet delay variation, and packet loss ratio,
and derives a service-specific \QoE{} KPI denoted as the interactivity score
(i-score). The evaluation of this methodology on a 5G non-public network is
reviewed in \autoref{sec:domains}.

\section{Related Work}
\label{sec:related}

\autoref{tab:survey-comp} summarizes the most relevant prior surveys. Four
broad lines of work exist. \autoref{sec:existing-qoe} reviews \QoE{} measurement
surveys. \autoref{sec:existing-domain} covers domain-specific surveys.
\autoref{sec:existing-ml} reviews ML-based prediction surveys.
\autoref{sec:existing-large} examines large-model and immersive-media surveys.
This survey addresses the gap stated in \autoref{sec:existing-gap}.

\subsection{\underline{\QoE{} Measurement Surveys}}
\label{sec:existing-qoe}

Kougioumtzidis \textit{et al.}~\cite{kougioumtzidis2022survey} survey ML-based \QoE{}
prediction for multimedia, covering supervised models and benchmarks, but treat
frameworks and ML as separate concerns. Barakabitze \textit{et al.}~\cite{j1-7} address
architecture and standardization for streaming \QoE{} management without
analyzing ML-framework coupling. Skorin-Kapov \textit{et al.}~\cite{j1-23} examine \QoE{}
concepts and challenges but predate federated and deep-learning adoption. Vega et
al.~\cite{j1-20} review predictive \QoE{} management in video streaming, covering
prediction models and adaptation strategies, but center on the model rather than
its coupling to measurement infrastructure. Torjemen and
Tabbane~\cite{torjemen2025survey} review \QoE{} standards (ITU, ETSI, 3GPP) and
5G/6G architectural enablers, and map \QoE{} requirements to network standards;
they organize prior surveys by the \QoE{} lifecycle (modeling, optimization, and
delivery), and do not provide an architectural taxonomy of how ML models are
coupled to measurement frameworks at the implementation level.

\subsection{\underline{Domain-Specific Surveys}}
\label{sec:existing-domain}

For adaptive streaming, Seufert \textit{et al.}~\cite{rw-6} and Barman and
Martini~\cite{p1-3} cover rebuffering and bitrate trade-offs but not the broader
ML toolchain. Alsader \textit{et al.}~\cite{alsader2025qoe} survey \QoE{}-driven adaptive
streaming toward 6G with SDN/NFV coverage, but remain streaming-specific: they
organize the field by streaming architectures and techniques rather than by how
measurement frameworks are coupled to ML models. Nasralla et
al.~\cite{nasralla2023exploring} address m-health \QoE{} in 6G with limited
coverage of general measurement tools or integration patterns. Skowronek et
al.~\cite{rw-3} survey \QoE{} in telemeetings and videoconferencing, covering
technical and contextual aspects but not the integration of measurement
frameworks with ML models. Garcia \textit{et al.}~\cite{garcia2026survey} review
privacy-preserving mechanisms (\textit{e.g.,} federated learning, differential privacy)
and their impact on \QoE{}; their primary lens is data protection, and while they
contrast network-centric and application-centric architectures, they do so to
explain where privacy mechanisms are placed, not how measurement frameworks are
coupled to ML models for \QoE{} prediction in general.

\subsection{\underline{ML-Based Prediction Surveys}}
\label{sec:existing-ml}

Ahmad \textit{et al.}~\cite{j2} compare supervised learning models for video streaming
\QoE{} but omit unsupervised, generative, and federated approaches. Huang et
al.~\cite{j1-18} focus on feature engineering without covering LLM-based or
RL-based paradigms. Laghari \textit{et al.}~\cite{rw-8} systematically analyze \QoE{}
frameworks for multimedia services but treat measurement and modeling as
separate concerns.

\subsection{\underline{Large-Model and Immersive-Media Surveys}}
\label{sec:existing-large}

A parallel line targets perceptual media quality directly. Zhang et
al.~\cite{zhang2025quality} survey quality assessment in the era of large
models. Ghosh and Singhal~\cite{ghosh2024review} review ML-based \QoE{} modeling
and adaptive streaming for conventional and 360-degree video. Wehner et
al.~\cite{wehner2026tutorial} provide an extensive tutorial on data-driven \QoE{}
modeling with explainable AI (XAI), covering explainability, context
adaptability, uncertainty quantification, and data decentralization through
federated learning. The tutorial organizes the landscape by \emph{explainability}
as the primary lens; it does not classify how measurement tools are
architecturally coupled to the ML models, nor does it map these coupling patterns
across different domains such as AR/VR or cloud gaming.

\subsection{\underline{The Integration Gap}}
\label{sec:existing-gap}

\autoref{tab:survey-comp} compares these surveys across seven coverage
dimensions: multimedia services, 6G, ML, immersive media, privacy, XAI and
generative AI. Prior surveys have extensively covered \QoE{} measurement, ML
prediction, and domain-specific constraints in isolation, but none classifies how
measurement frameworks and ML models are coupled, or the enhancement mechanisms
that such coupling enables. Torjemen and Tabbane~\cite{torjemen2025survey} focus
on network standards, Wehner \textit{et al.}~\cite{wehner2026tutorial} focus on
explainability, and Garcia \textit{et al.}~\cite{garcia2026survey} focus on privacy. The
field therefore lacks a taxonomy that explains how measurement infrastructure and
ML models are jointly designed and deployed. This survey addresses this gap; its
research questions and contributions are stated in \autoref{sec:intro}, and
\autoref{sec:method} defines the set of works and the coding rule that answering
them requires.

\begin{table*}[]
\centering
\caption{{Comparison of existing \QoE{} surveys across various domains. Rows are
ordered by publication year, and the final row is this survey. A check mark
(\checkmark) indicates that the survey addresses the corresponding dimension,
while a cross (\ding{55}) indicates that the dimension is not addressed.
\textbf{Column abbreviations:} \textbf{MM} = Multimedia Streaming;
\textbf{6G} = 5G/6G Networking; \textbf{ML} = Machine Learning Prediction;
\textbf{Immer.} = Immersive Media (AR/VR/Cloud Gaming); \textbf{Privacy} = Data
Privacy and Security; \textbf{XAI} = Explainable AI; \textbf{GenAI} = Generative
AI and LLMs.}}
\label{tab:survey-comp}
\renewcommand{\arraystretch}{1.3}
\setlength{\tabcolsep}{4pt}
\scriptsize
\rowcolors{2}{gray!10}{white}
\begin{tabular}{@{}>{\raggedright\arraybackslash}m{1.5cm}
                  >{\raggedright\arraybackslash}m{2.3cm}
                  >{\centering\arraybackslash}m{0.6cm} 
                  >{\centering\arraybackslash}m{0.6cm} 
                  >{\centering\arraybackslash}m{0.6cm} 
                  >{\centering\arraybackslash}m{1cm} 
                  >{\centering\arraybackslash}m{0.9cm} 
                  >{\centering\arraybackslash}m{0.6cm} 
                  >{\centering\arraybackslash}m{0.6cm} 
                  >{\raggedright\arraybackslash}m{6cm}@{}}
\toprule
\rowcolor{headerblue}
{\color{white}\textbf{Ref. (Year)}} &
{\color{white}\textbf{Focus Area}} &
{\color{white}\textbf{MM}} &
{\color{white}\textbf{6G}} &
{\color{white}\textbf{ML}} &
{\color{white}\textbf{Immer.}} &
{\color{white}\textbf{Privacy}} &
{\color{white}\textbf{XAI}} &
{\color{white}\textbf{GenAI}} &
{\color{white}\textbf{Key Contribution}} \\
\midrule
\cite{rw-8} (2017) & \QoE{} frameworks & \textcolor{blue}{\checkmark} & \textcolor{red}{\ding{55}} & \textcolor{red}{\ding{55}} & \textcolor{red}{\ding{55}} & \textcolor{red}{\ding{55}} & \textcolor{red}{\ding{55}} & \textcolor{red}{\ding{55}} & Analyzes \QoE{} frameworks. \\
\cite{j1-20} (2018) & Predictive \QoE{} & \textcolor{blue}{\checkmark} & \textcolor{red}{\ding{55}} & \textcolor{blue}{\checkmark} & \textcolor{red}{\ding{55}} & \textcolor{red}{\ding{55}} & \textcolor{red}{\ding{55}} & \textcolor{red}{\ding{55}} & Reviews predictive \QoE{}. \\
\cite{rw-4} (2019) & 5G technology & \textcolor{red}{\ding{55}} & \textcolor{blue}{\checkmark} & \textcolor{red}{\ding{55}} & \textcolor{red}{\ding{55}} & \textcolor{red}{\ding{55}} & \textcolor{red}{\ding{55}} & \textcolor{red}{\ding{55}} & Reviews QoS/\QoE{} in 5G. \\
\cite{rw-9} (2019) & Unified comms & \textcolor{blue}{\checkmark} & \textcolor{red}{\ding{55}} & \textcolor{red}{\ding{55}} & \textcolor{red}{\ding{55}} & \textcolor{red}{\ding{55}} & \textcolor{red}{\ding{55}} & \textcolor{red}{\ding{55}} & Surveys \QoE{} for comms. \\
\cite{kougioumtzidis2022survey} (2022) & Multimedia \QoE{} & \textcolor{blue}{\checkmark} & \textcolor{red}{\ding{55}} & \textcolor{blue}{\checkmark} & \textcolor{blue}{\checkmark} & \textcolor{red}{\ding{55}} & \textcolor{red}{\ding{55}} & \textcolor{red}{\ding{55}} & Surveys ML-based \QoE{}. \\
\cite{rw-3} (2022) & Telemeetings & \textcolor{blue}{\checkmark} & \textcolor{red}{\ding{55}} & \textcolor{red}{\ding{55}} & \textcolor{blue}{\checkmark} & \textcolor{red}{\ding{55}} & \textcolor{red}{\ding{55}} & \textcolor{red}{\ding{55}} & Surveys telemeetings. \\
\cite{nasralla2023exploring} (2023) & 6G m-health & \textcolor{blue}{\checkmark} & \textcolor{blue}{\checkmark} & \textcolor{red}{\ding{55}} & \textcolor{red}{\ding{55}} & \textcolor{red}{\ding{55}} & \textcolor{red}{\ding{55}} & \textcolor{red}{\ding{55}} & Explores 6G m-health. \\
\cite{alsader2025qoe} (2025) & 6G streaming & \textcolor{blue}{\checkmark} & \textcolor{blue}{\checkmark} & \textcolor{red}{\ding{55}} & \textcolor{blue}{\checkmark} & \textcolor{red}{\ding{55}} & \textcolor{red}{\ding{55}} & \textcolor{red}{\ding{55}} & Surveys 6G streaming. \\
\cite{ghosh2024review} (2025) & ML \QoE{} & \textcolor{blue}{\checkmark} & \textcolor{red}{\ding{55}} & \textcolor{blue}{\checkmark} & \textcolor{red}{\ding{55}} & \textcolor{red}{\ding{55}} & \textcolor{red}{\ding{55}} & \textcolor{red}{\ding{55}} & Reviews ML \QoE{}. \\
\cite{panahi2025bridging} (2025) & Dual-layer \QoE{} & \textcolor{blue}{\checkmark} & \textcolor{red}{\ding{55}} & \textcolor{blue}{\checkmark} & \textcolor{red}{\ding{55}} & \textcolor{red}{\ding{55}} & \textcolor{red}{\ding{55}} & \textcolor{red}{\ding{55}} & Integrates network and LLM. \\
\cite{zhang2025quality} (2025) & Large-model quality & \textcolor{blue}{\checkmark} & \textcolor{red}{\ding{55}} & \textcolor{blue}{\checkmark} & \textcolor{red}{\ding{55}} & \textcolor{red}{\ding{55}} & \textcolor{red}{\ding{55}} & \textcolor{blue}{\checkmark} & Surveys large-model quality. \\
\cite{torjemen2025survey} (2025) & 5G/6G standards & \textcolor{blue}{\checkmark} & \textcolor{blue}{\checkmark} & \textcolor{blue}{\checkmark} & \textcolor{blue}{\checkmark} & \textcolor{red}{\ding{55}} & \textcolor{red}{\ding{55}} & \textcolor{red}{\ding{55}} & Reviews standards. \\
\cite{wehner2026tutorial} (2026) & XAI for \QoE{} & \textcolor{blue}{\checkmark} & \textcolor{red}{\ding{55}} & \textcolor{blue}{\checkmark} & \textcolor{red}{\ding{55}} & \textcolor{blue}{\checkmark} & \textcolor{blue}{\checkmark} & \textcolor{red}{\ding{55}} & Tutorial on XAI. \\
\cite{garcia2026survey} (2026) & Privacy \QoE{} & \textcolor{blue}{\checkmark} & \textcolor{blue}{\checkmark} & \textcolor{blue}{\checkmark} & \textcolor{blue}{\checkmark} & \textcolor{blue}{\checkmark} & \textcolor{red}{\ding{55}} & \textcolor{red}{\ding{55}} & Reviews privacy. \\
\textbf{This survey} & \textbf{Framework-ML int.} & \textbf{\textcolor{blue}{\checkmark}} & \textbf{\textcolor{blue}{\checkmark}} & \textbf{\textcolor{blue}{\checkmark}} & \textbf{\textcolor{blue}{\checkmark}} & \textbf{\textcolor{blue}{\checkmark}} & \textbf{\textcolor{blue}{\checkmark}} & \textbf{\textcolor{blue}{\checkmark}} & \textbf{First unified study of integration across all dimensions.} \\
\bottomrule
\end{tabular}
\vspace{-1.5em}
\end{table*}

\section{Survey Methodology}
\label{sec:method}

This section describes how the reviewed works were found, selected and coded, so
that the answers to \texttt{RQ1}--\texttt{RQ4} rest on a defined set of works.
\autoref{sec:method-search} gives the sources and the search rounds.
\autoref{sec:method-selection} gives the selection rule and the count at each
step. \autoref{sec:method-coding} defines the coding rule that orders
\autoref{sec:landscape} to \autoref{sec:challenge}. The three parts follow the
order in which the set of works was built.

\subsection{\underline{Search Strategy}}
\label{sec:method-search}

The search used four databases: Google Scholar, IEEE Xplore, ResearchGate and
ScienceDirect. In the first round, the websites of crowdsourcing platforms and
of closed-source measurement products were also searched, because such tools are
often described only in vendor documents. Each round ran two passes. The first
pass combined \QoE{} terms with general ML terms. The second pass combined \QoE{}
terms with the names of ML families that the first pass had not returned.
\autoref{tab:search-terms} lists the terms. Terms within a group were joined by
OR. The \QoE{} group was joined by AND with one other group.

\begin{table}[]
\centering
\caption{{Search terms. Each query joined the \QoE{} group by AND with one other
group.}}
\label{tab:search-terms}
\renewcommand{\arraystretch}{1.3}
\setlength{\tabcolsep}{4pt}
\footnotesize
\begin{tabular}{@{}>{\raggedright\arraybackslash}m{2.6cm}>{\raggedright\arraybackslash}m{5.2cm}@{}}
\toprule
\rowcolor{headerblue}
{\color{white}\textbf{Group}} & {\color{white}\textbf{Terms}} \\
\midrule
\QoE{} & "quality of experience", QoE, "mean opinion score", MOS \\
Tools (round 1) & framework, tool, platform, testbed, crowdsourcing, "open source" \\
ML, first pass & "machine learning", "deep learning" \\
ML, second pass & "reinforcement learning", "federated learning", unsupervised, "generative adversarial", transformer, "large language model" \\
\bottomrule
\end{tabular}
\vspace{-0.5em}
\end{table}

The search ran in three rounds, listed in \autoref{tab:search-rounds}. The third
round was run by a co-author with the same sources and the same two passes.
Duplicates were removed across sources and across rounds.

\begin{table}[]
\centering
\caption{{Records per search round. The four databases searched in every round
were Google Scholar, IEEE Xplore, ResearchGate and ScienceDirect. ``Unique''
counts records after duplicates are removed both within a round and against all
earlier rounds.}}
\label{tab:search-rounds}
\renewcommand{\arraystretch}{1.3}
\setlength{\tabcolsep}{4pt}
\footnotesize
\begin{tabular}{@{}lcccc@{}}
\toprule
\rowcolor{headerblue}
{\color{white}\textbf{Round (year)}} &
{\color{white}\textbf{Found}} &
{\color{white}\textbf{Duplicates}} &
{\color{white}\textbf{Unique}} &
{\color{white}\textbf{Kept}} \\
\midrule
1 (2024) & 92 & 19 & 73 & 41 \\
2 (2025) & 34 & 6 & 28 & 11 \\
3 (2026) & 81 & 24 & 57 & 27 \\
\midrule
\textbf{Total} & \textbf{207} & \textbf{49} & \textbf{158} & \textbf{79} \\
\bottomrule
\end{tabular}
\vspace{-0.5em}
\end{table}

\subsection{\underline{Selection and Counts}}
\label{sec:method-selection}

A record was kept when it measures, predicts, or controls the \QoE{} of a
multimedia service through a measurement framework, an ML model, or both, or when
it surveys this area. The three rounds returned 207 records
(\autoref{tab:search-rounds}). After duplicate removal across sources and across
rounds, 158 unique records remained. Title and abstract screening then excluded
79 of them, and the remaining 79 records were read in full.

Full-text reading kept 64 primary works and 15 surveys. Four further works were
added by hand from the reference lists of kept works: three adaptive-bitrate
controllers~\cite{mao2017pensieve,huang2019comyco,akhtar2018oboe} and one
industry white paper~\cite{sr-2}. Four surveys were also added by hand from the
same reference lists, because they were cited repeatedly by kept works but had
not appeared in the database queries. One open-source framework appears in the
log as two records, its paper~\cite{mine-archive} and its code
repository~\cite{f4-8}, and is counted as one work.

\begin{figure}[]
    \centering
    \includegraphics[width=0.8\columnwidth]{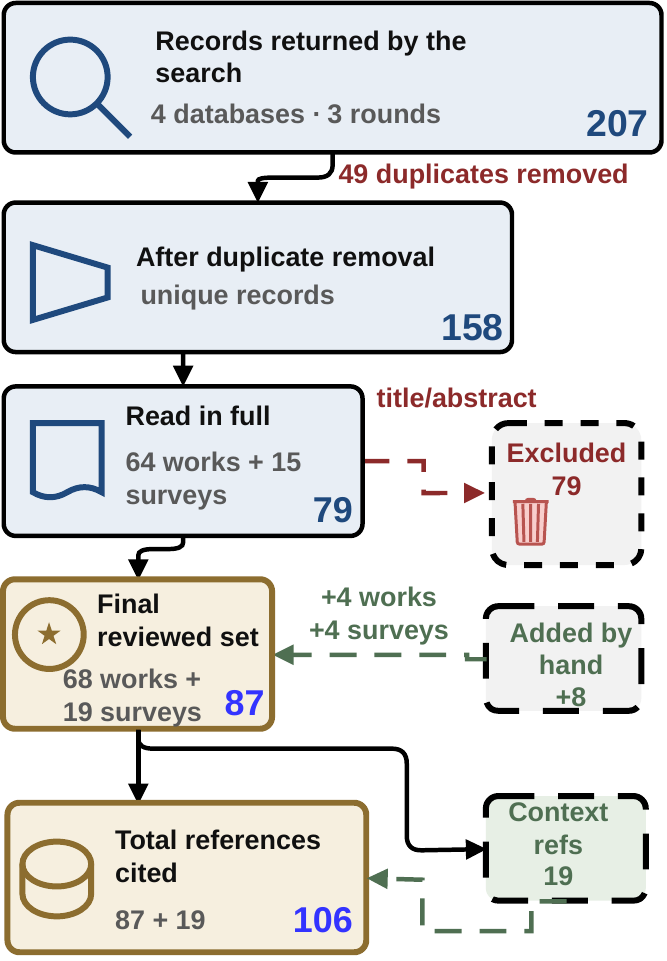}
    \caption{Selection funnel from the initial search to the 106 references cited
    in this survey. The search returned 207 records; 49 duplicates were removed
    and 79 were excluded at title/abstract screening. Full-text reading kept 64
    primary works and 15 surveys, and four works and four surveys were added by
    hand. The 87 reviewed records, together with 19 references that supply
    definitions, standards and context, give the 106 references cited.}
    \label{fig:selection-flow}
\vspace{-1em}
\end{figure}

The final set therefore holds 68 primary works and 19 surveys, and these 87
records are the ones analyzed in \autoref{sec:landscape} to
\autoref{sec:challenge} and compared in \autoref{sec:related}. The manuscript
cites 106 references in total: the 87 works and surveys, plus 19 further
references that supply definitions, standards and context in
\autoref{sec:intro} and \autoref{sec:background} and are not part of the
reviewed set. \autoref{fig:selection-flow} shows the count at each step of the
pipeline.

The measurement tools reviewed in \autoref{sec:landscape} were further selected
on three criteria. Relevance: direct applicability to \QoE{} measurement in
multimedia services. Impact: adoption in academic literature and industry,
evidenced by citation count and deployment. Diversity: representation of distinct
technical architectures (active and passive) and licensing models (open and
closed source).

\subsection{\underline{Coding Scheme}}
\label{sec:method-coding}

Each reviewed work receives one primary code and several tags. The primary code
is exclusive. It is decided by where model inference runs relative to the
measurement loop:

\begin{itemize}
  \item \emph{Measurement only (M):} the work uses no learned model. Tools,
        testbeds, datasets, standards and fixed formulas such as
        \texttt{ITU-T P.1203} fall here. Each M work is also marked with the
        coupling patterns that its data output \emph{can support}.
  \item \emph{Loose coupling (L):} a model is trained and tested offline on
        stored measurement data.
  \item \emph{Tight coupling (T):} model inference runs inside the live
        measurement or control loop, on the client, the edge, or the network.
  \item \emph{Distributed/federated (D):} training or decisions are spread over
        several devices or network nodes that share model updates or decisions.
\end{itemize}

A work is coded by what it built and evaluated, not by what it states could be
done later. A prediction model tested only on a dataset is therefore coded L,
even when its authors describe it as real-time. Surveys, position papers and
optimizers without a learned model receive no pattern code. The tags record the
learning family, the framework type (closed or open source), the service domain,
the action taken on the service, if any, and the evidence level (simulation,
laboratory or testbed, field data, or production deployment). A class supported
by a single work is recorded as an emerging direction rather than as a category.
Each code was checked against the full text of the work where it was available
(56 of the 68 works) and otherwise against its abstract or a companion paper by
the same authors.

The results of the coding are reported in \autoref{sec:patterns}. Before that,
\autoref{sec:landscape} reviews the measurement frameworks and ML models to which
the coding applies.

\section{Measurement Frameworks and ML Models}
\label{sec:landscape}

This section reviews the two components that every integrated system joins: the
measurement frameworks of stage~2 in \autoref{fig:all-in-one}, which collect user
experience and telemetry, and the ML models of stage~3, which learn from these
data. It follows the path of the data. \autoref{sec:tools-workflow} gives the
measurement workflow that every framework implements.
\autoref{sec:tools-closed} and \autoref{sec:tools-open} review closed- and
open-source frameworks, which implement the same workflow but differ in
licensing and replicability. \autoref{sec:ml-prediction} reviews the ML models
by learning paradigm. Each work in
Tables~\ref{tab:closed-source}--\ref{tab:summl-rl} carries the code defined in
\autoref{sec:method-coding}; how the two components are coupled is analyzed in
\autoref{sec:patterns}. The selection criteria for these tools are given in
\autoref{sec:method-selection}.

\subsection{\underline{QoE Measurement Workflow}}
\label{sec:tools-workflow}

\QoE{} measurement tools have to record user-perceived parameters automatically,
such as video quality and stalls, for ML model training. A tool has to meet four
requirements: it has to simulate user behavior, record the parameters
automatically, calculate scores such as the mean opinion score (MOS), and
validate the scores against subjective data.

\begin{figure}
    \centering
    \includegraphics[width=0.9\linewidth,clip,trim={0.2cm 0.5cm 2.5cm 0cm}]{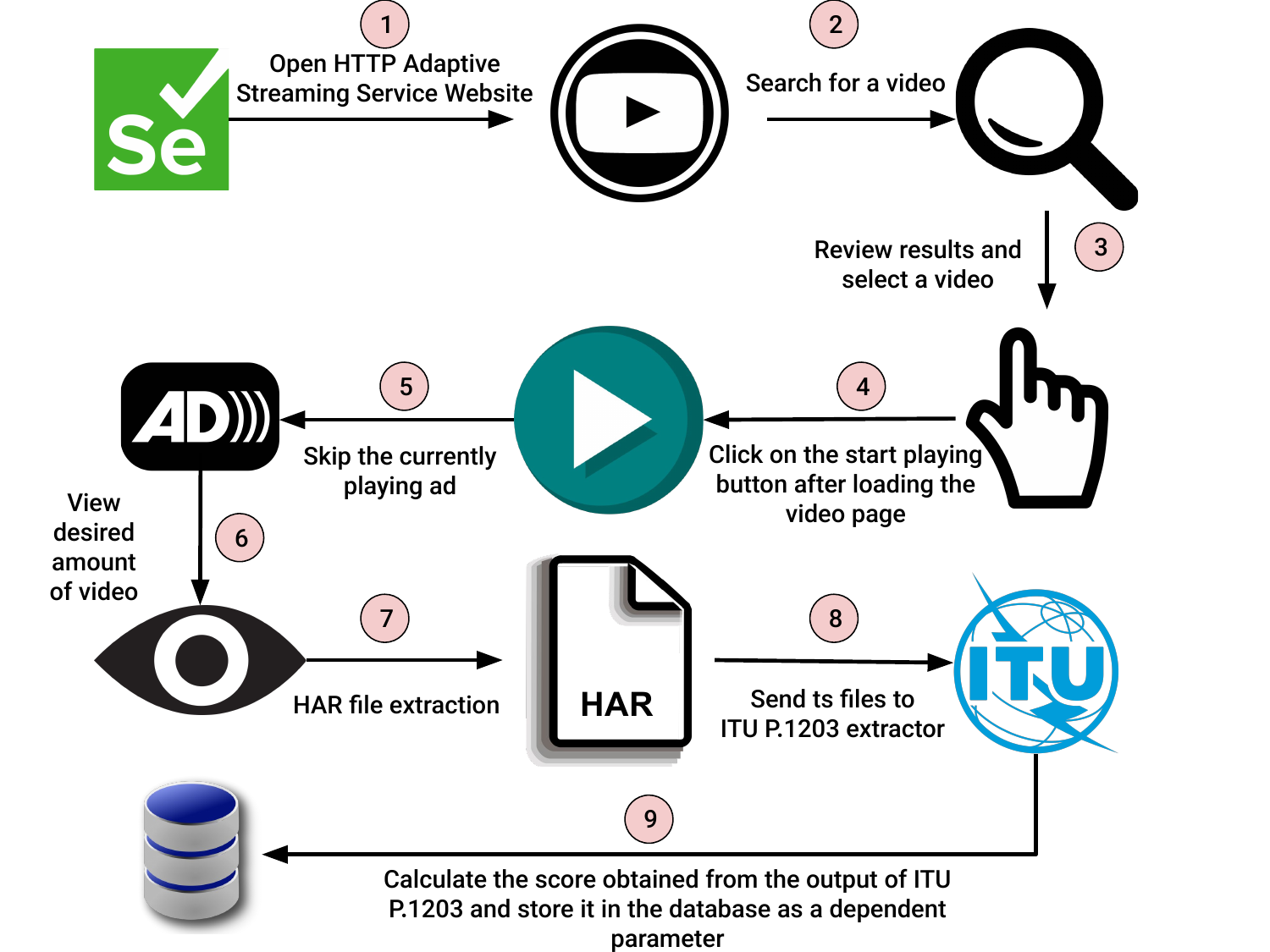}
    \caption{Seven-step \texttt{ITU-T P.1203} workflow for measuring
    video-streaming \QoE{}, from opening the service to storing labeled data for
    ML training.}
    \label{fig:final-examp}
\vspace{-1em}
\end{figure}

\autoref{fig:final-examp} shows the \texttt{ITU-T P.1203}
workflow~\cite{mine-archive,f4-8}: (1) open an HTTP adaptive streaming (HAS)
service, (2) search and select a video, (3) play the video, (4) extract an HTTP
Archive (HAR) file, (5) isolate the .ts files, (6) calculate quality per file,
and (7) store labeled data for ML training. This workflow enables distributed
data collection across user devices.

\subsection{\underline{Closed-Source Platforms}}
\label{sec:tools-closed}

A \emph{closed-source platform} is any \QoE{} measurement tool, framework,
algorithm, or testbed whose implementation is not publicly replicable. This
includes commercial solutions, operator-internal systems, proprietary
algorithms that require non-public production data, and controlled testbed
studies. The shared property is non-replicability, not artifact form. The CFA
algorithm~\cite{sr-3}, for example, is included despite appearing in academic
publications because replication requires proprietary CDN data. Vendor
documents~\cite{sr-4} and cloud gaming studies~\cite{sr-7} are included because
the evaluated platform is itself a closed commercial service. This definition
covers the non-open landscape of \QoE{} measurement.

\ignore{
\begin{figure}
    \centering
    \includegraphics[width=0.95\linewidth,clip,trim={0.2cm 0.75cm 0.5cm 0.5cm}]{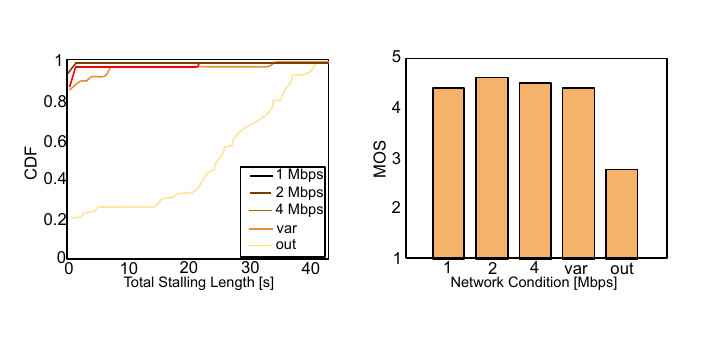}
    \caption{Mean opinion score analysis of video-playback stalls under varying
    network conditions~\cite{sr-1}.}
    \label{fig:art-1}
\vspace{-1em}
\end{figure}
}

\begin{table*}[]
\centering
\caption{{Closed-source \QoE{} measurement frameworks: application,
data-collection methods, and key features. Superscripts give the code of each
work (\autoref{sec:method-coding}).}}
\label{tab:closed-source}
\renewcommand{\arraystretch}{1.3}
\setlength{\tabcolsep}{4pt}
\footnotesize
\rowcolors{2}{gray!10}{white}
\begin{tabular}{@{}>{\raggedright\arraybackslash}m{0.8cm}
                  >{\centering\arraybackslash}m{1cm}
                  >{\raggedright\arraybackslash}m{4.2cm}
                  >{\raggedright\arraybackslash}m{5cm}
                  >{\raggedright\arraybackslash}m{4.7cm}@{}}
\toprule
\rowcolor{headerblue}
{\color{white}\textbf{Ref.}} & {\color{white}\textbf{Year}} &
{\color{white}\textbf{Application}} &
{\color{white}\textbf{Data Collection}} &
{\color{white}\textbf{Features}} \\
\midrule
\cite{sr-8}\textsuperscript{M}  & 2011 & IPTV networks & Objective experiments on a network platform & Measures network parameters and their impact on \QoE{}. \\
\cite{sr-11}\textsuperscript{L} & 2012 & Mobile network optimization & Subjective tests with software agents collecting network performance & Combines user \QoE{} with technical parameters; reports high model accuracy. \\
\cite{sr-1}\textsuperscript{M}  & 2015 & \QoE{}-based traffic optimization & YoMoApp simulation of YouTube behavior & Passive client-side measurement. \\
\cite{sr-12}\textsuperscript{M} & 2015 & YouTube \QoE{} on smartphones & YoMoApp measurement plus a lab subjective study (52 participants, WiFi with emulated bandwidth) & Accurate monitoring of player state and buffer; no model is trained. \\
\cite{sr-3}\textsuperscript{T}  & 2016 & CFA algorithm & Real user data with \QoE{} measurements and session features & Reported high prediction accuracy and scalability; pilot deployment for CDN and bitrate selection. \\
\cite{sr-17}\textsuperscript{L} & 2017 & In-smartphone traffic measurement & Field trial on operational cellular networks & Combines objective and subjective measurements. \\
\cite{sr-19}\textsuperscript{M} & 2019 & IoT service \QoE{} & MOS survey of institutional IoT users (ACR-HR scale) & Five-step framework using real data. \\
\cite{sr-20}\textsuperscript{L} & 2019 & Proactive LTE management & Cell-level LTE data over five weeks & Proactive network management framework. \\
\cite{sr-2}\textsuperscript{M}  & 2023 & Video-stream quality detection & Appium, Selenium, and IR tools to detect displayed frames & Automation across devices; visualization. \\
\cite{sr-4}\textsuperscript{T}  & 2023 & Video quality by location and device & Sandvine traffic analytics for video KPIs & Classifies encrypted videos and reports quality. \\
\cite{sr-5}\textsuperscript{M}  & 2023 & Mobile app and device \QoE{} & Network lab evaluation of mobile apps & High control and repeatability. \\
\cite{sr-7}\textsuperscript{M}  & 2023 & Mobile cloud gaming \QoE{} & Lab subjective tests: 22 players, 23 network conditions (NetEm, ALTRUIST) & CS:GO over Steam Remote Play; MOS falls to 1.4 at 12\% packet loss. \\
\bottomrule
\end{tabular}
\vspace{-1.5em}
\end{table*}

\autoref{tab:closed-source} summarizes closed-source tools. Key frameworks
include IPTV \QoE{} analysis~\cite{sr-8}, mobile network optimization combining
\QoE{} and QoS~\cite{sr-11}, and YoMoApp for YouTube \QoE{}
measurement~\cite{sr-1,sr-12}, validated against subjective evaluations.

\ignore{
\begin{figure}
    \centering
    \includegraphics[width=0.95\linewidth,clip,trim={0.5cm 0.5cm 0.5cm 0cm}]{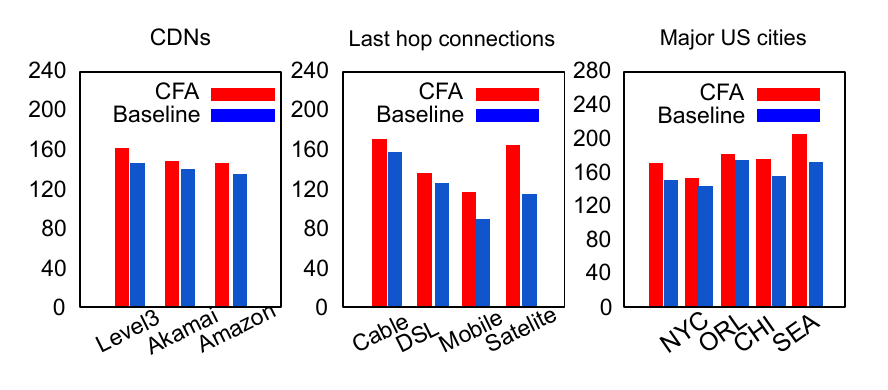}
    \caption{CFA algorithm results by content delivery network, last-hop
    connection type, and city~\cite{sr-3}.}
    \label{fig:art-3}
\vspace{-1em}
\end{figure}
}

The CFA algorithm~\cite{sr-3} reports high accuracy and
scalability.
ML-based models~\cite{sr-17} report above
90\% accuracy from smartphone measurements. The IoT \QoE{}
framework~\cite{sr-19} has five steps, one of which is a MOS survey. LTE network
management~\cite{sr-20} employs ML for proactive \QoE{} prediction, trained and
tested offline on five weeks of cell-level data. \QoE{} analysis
systems~\cite{sr-2} examine encoding, network, and device factors. Laboratory
frameworks~\cite{sr-5,sr-7} enable controlled \QoE{} evaluation. In the cloud
gaming study~\cite{sr-7}, round-trip times from 2 to 170~ms keep the MOS near or
above 3.0; a 200~ms delay burst every 15~s still scores 4.5, and 12\% packet
loss lowers the MOS to 1.4.

Closed-source solutions typically employ probe-based active measurement
mechanisms deployed at network edges or user devices. They prioritize
high-fidelity data collection for specific proprietary applications, such as
IPTV and cellular optimization, but often lack the flexibility for cross-domain
adaptation. Most of them store their data for later analysis. Only two run a
model inside the live loop, the CFA system~\cite{sr-3} and the Sandvine
analytics product~\cite{sr-4}, as \autoref{sec:patterns} details.

\subsection{\underline{Open-Source Platforms}}
\label{sec:tools-open}

\begin{table*}[]
\centering
\caption{{Open-source \QoE{} measurement frameworks: application,
data-collection methods, and key features. Superscripts give the code of each
work (\autoref{sec:method-coding}).}}
\label{tab:open-source}
\renewcommand{\arraystretch}{1.3}
\setlength{\tabcolsep}{4pt}
\footnotesize
\rowcolors{2}{gray!10}{white}
\begin{tabular}{@{}>{\raggedright\arraybackslash}m{0.8cm}
                  >{\centering\arraybackslash}m{1cm}
                  >{\raggedright\arraybackslash}m{4.2cm}
                  >{\raggedright\arraybackslash}m{5cm}
                  >{\raggedright\arraybackslash}m{4.7cm}@{}}
\toprule
\rowcolor{headerblue}
{\color{white}\textbf{Ref.}} & {\color{white}\textbf{Year}} &
{\color{white}\textbf{Application}} &
{\color{white}\textbf{Data Collection}} &
{\color{white}\textbf{Features (Language)}} \\
\midrule
\cite{sr-10}\textsuperscript{M} & 2014 & Android YouTube \QoE{} app & Pilot experiment with 17 users & YouTube \QoE{} on Android (Android SDK). \\
\cite{sr-14}\textsuperscript{M} & 2015 & Smartphone video \QoE{} & Objective experiments with a software tool & Records \QoE{} parameters and models stalls. \\
\cite{sr-16}\textsuperscript{M} & 2015 & Mobile broadband \QoE{} & Data from three mobile operators & Real data and reproducible results. \\
\cite{sr-15}\textsuperscript{M} & 2019 & IoT multimedia \QoE{} & Simulation with randomly generated data & \QoE{}-based resource estimation (Java). \\
\cite{sr-21}\textsuperscript{M} & 2022 & DASH adaptive streaming & Emulation of joint edge-caching and recommendation policies, built on goDASH & Emulates multi-user streaming over MEC and cloud nodes. \\
\cite{sr-6}\textsuperscript{M}  & 2023 & Crowdsourced \QoE{} evaluation & Crowdsourcing framework with simulation on real data & Reduces cost (by 32--46\%) and measurement delay (by 51--69\%). \\
\cite{sr-9}\textsuperscript{M}  & 2023 & Containerized video quality & Objective experiments on video samples & 14 video quality metrics (Python, Bash). \\
\cite{sr-18}\textsuperscript{M} & 2024 & Longitudinal \QoE{} study & Daily Video app simulating real-world consumption & Studies user behavior over time (Python, JavaScript, React Native). \\
\cite{mine-archive,f4-8}\textsuperscript{L} & 2025 & ML-based \QoE{} parameter extraction & Selenium and \texttt{ITU-T P.1203} in simulated and real modes & Predicts the P.1203 MOS from five network parameters (RF, $R^2$ 0.968) (Python, Docker). \\
\bottomrule
\end{tabular}
\vspace{-0.5em}
\end{table*}

\begin{figure}
    \centering
    \includegraphics[width=0.95\linewidth,clip,trim={0.1cm 0.1cm 0.1cm 0.1cm}]{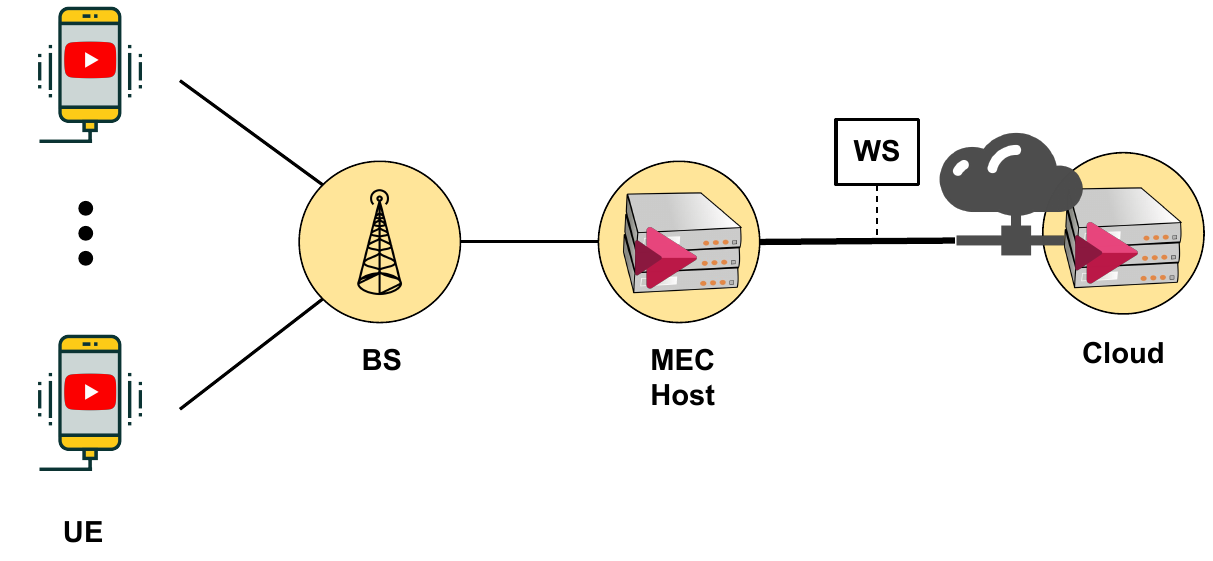}
    \vspace{-1em}
    \caption{QoE-DASH architecture, from the user-equipment layer to the cloud
    server~\cite{sr-21}.}
    \label{fig:article-17}
\vspace{-1.5em}
\end{figure}

\autoref{tab:open-source} summarizes open-source \QoE{} tools. Key tools include
Android apps for YouTube \QoE{}~\cite{sr-10}, VLQoE for smartphone video
evaluation~\cite{sr-14}, mobile broadband \QoE{} frameworks~\cite{sr-16}, IoT
\QoE{} models~\cite{sr-15}, and \QoE{}-DASH for DASH streaming
simulation~\cite{sr-21}. The \QoE{}-DASH architecture~(\autoref{fig:article-17})
comprises user equipment (UE), a base station (BS), an MEC host that serves as
an edge cache, and a cloud server that stores content.

\begin{figure}
    \centering
    \includegraphics[width=0.95\linewidth,clip,trim={1cm 0.75cm 0.5cm 1cm}]{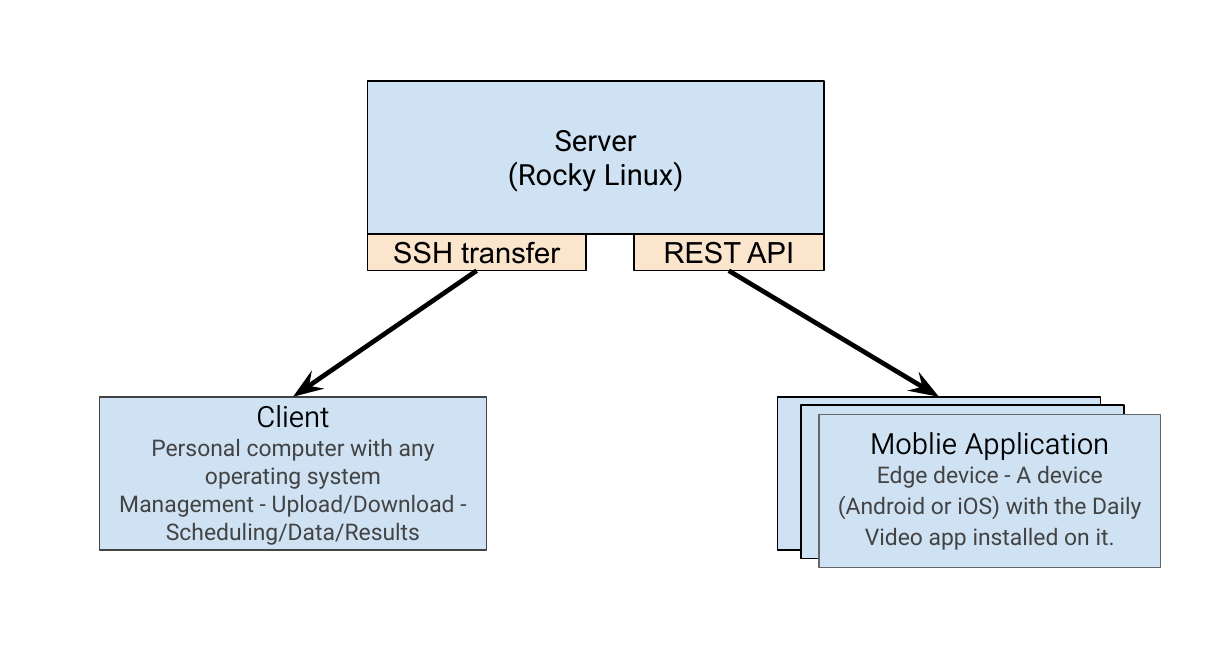}
    \vspace{-1em}
    \caption{Three-layer architecture of the Daily Video tool: server, client,
    and user mobile-app layers.}
    \label{fig:3-layer}
\vspace{-1em}
\end{figure}

\begin{figure*}
    \centering
    \includegraphics[width=0.85\linewidth]{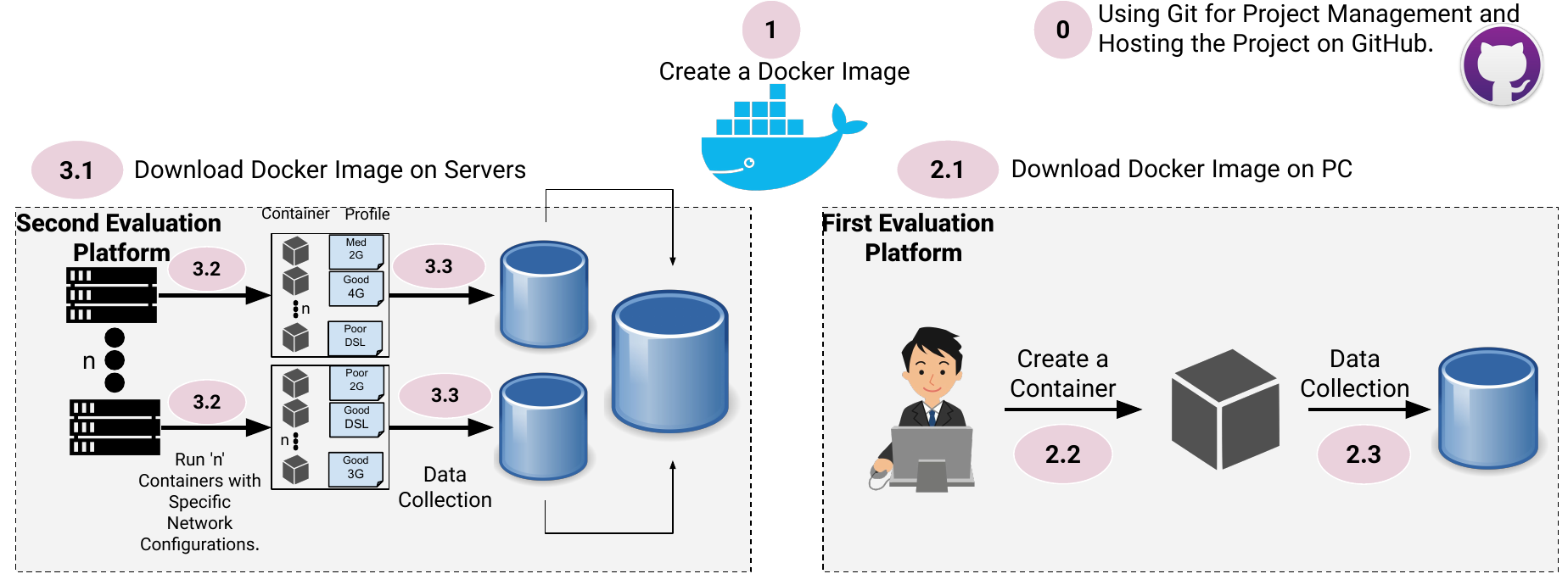}
    \caption{QoE measurement framework in which user PCs connect to servers
    that emulate distinct network profiles inside Docker
    containers~\cite{mine-archive,f4-8}.}
    \label{fig:impl}
\vspace{-1em}
\end{figure*}

\QoE{}-DASH builds on the goDASH player and runs the servers in Docker
containers. The Daily Video app~\cite{sr-18} enables longitudinal \QoE{} studies
with a three-layer architecture~(\autoref{fig:3-layer}): a server for storage and
exchange, a client for experiment management, and a mobile app built with React
Native.

\begin{figure}
    \centering
    \includegraphics[width=0.95\linewidth,clip,trim={0cm 1.5cm 0cm 0.5cm}]{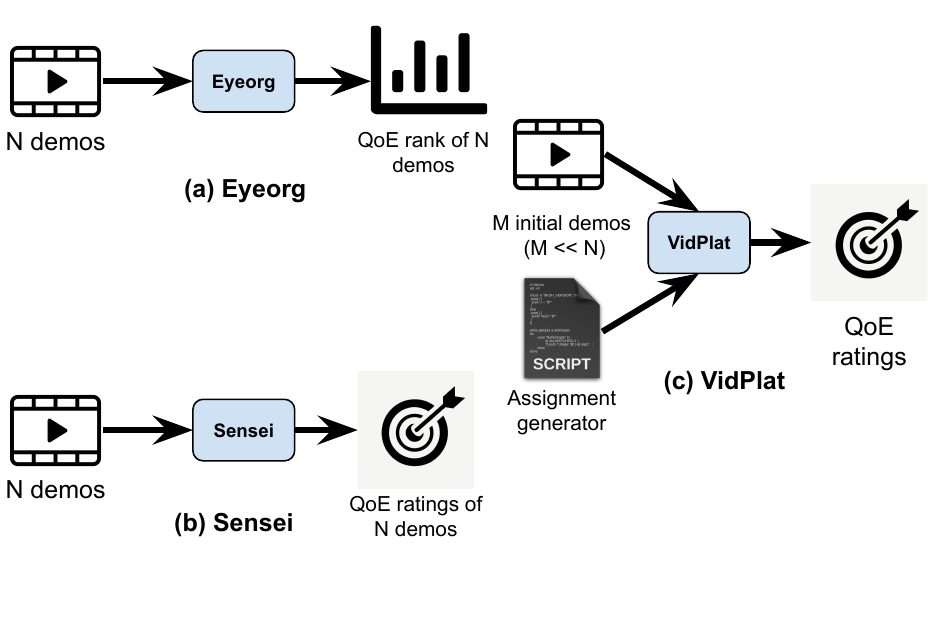}
    \caption{Comparison of VidPlat's dynamic crowdsourcing workflow against
    earlier \QoE{} tools that require predefined samples and rating
    counts~\cite{sr-6}.}
    \label{fig:vidplat-new}
\vspace{-1em}
\end{figure}

VidPlat~\cite{sr-6} uses dynamic generator logic to determine samples from user
feedback~(\autoref{fig:vidplat-new}), which removes the need to prespecify
samples and rating counts. The containerized video quality tool~\cite{sr-9}
computes 14 metrics on 4K video; one second of video takes from 1.1~s (SSIM) to
1,813~s (VIIDEO), and therefore only some of the metrics suit live streaming.

The framework of~\cite{mine-archive,f4-8} uses Selenium and \texttt{ITU-T P.1203}
for automated \QoE{} measurement~(\autoref{fig:impl}). It uses simplified network
parameters: it labels about 22,000 video segments with \texttt{ITU-T P.1203} and
trains ML models that predict the MOS from five network parameters.

Open-source frameworks increasingly use containerization with Docker and
client-side scripting in Python and JavaScript to enable flexible, distributed
data collection. Apart from the framework of~\cite{mine-archive,f4-8}, none of
the open-source tools trains a model itself. Their stored outputs can support
loose coupling. The emulated edge nodes of \QoE{}-DASH could host a learned
policy (\autoref{sec:patterns}). Open code also allows broader community
contribution and adaptation to different network environments.

\subsection{\underline{ML Models by Learning Paradigm}}
\label{sec:ml-prediction}

\begin{table*}[]
\centering
\caption{{Supervised and unsupervised ML models for \QoE{} prediction: core
concepts and reported results. Superscripts give the code of each work
(\autoref{sec:method-coding}).}}
\label{tab:summl}
\renewcommand{\arraystretch}{1.2}
\setlength{\tabcolsep}{3pt}
\footnotesize
\rowcolors{2}{gray!10}{white}
\begin{tabular}{@{}>{\raggedright\arraybackslash}m{0.6cm}
                  >{\centering\arraybackslash}m{0.6cm}
                  >{\raggedright\arraybackslash}m{2cm}
                  >{\raggedright\arraybackslash}m{4.4cm}
                  >{\raggedright\arraybackslash}m{4.2cm}
                  >{\raggedright\arraybackslash}m{5cm}@{}}
\toprule
\rowcolor{headerblue}
{\color{white}\textbf{Ref.}} & {\color{white}\textbf{Year}} &
{\color{white}\textbf{Algorithm}} & {\color{white}\textbf{Advantage}} &
{\color{white}\textbf{Key Idea}} & {\color{white}\textbf{Results}} \\
\midrule
\cite{ml-18}\textsuperscript{L} & 2020 & Improved TCN & Long-term dependencies; 3$\times$ faster than LSTM & Dilated causal convolutions & PCC 0.82--0.89 on three databases. \\
\cite{Hamidi}\textsuperscript{L} & 2025 & Transformer & Robust cross-device estimation & Self-attention over streaming KPIs & Beats \texttt{ITU-T P.1203} on RMSE, PCC, SCC. \\
\cite{ml-21}\textsuperscript{L} & 2019 & BLSTM + LSTM & Strong temporal modeling & Temporal relationships for continuous \QoE{} & LCC 0.894, SROCC 0.830. \\
\cite{ml-6}\textsuperscript{L} & 2019 & Stacked LSTM & Quality, rebuffering and recency & Quality-metric and playback-state inputs & LCC up to 0.947 on four databases. \\
\cite{ml-8}\textsuperscript{L} & 2018 & 3D CNN + SVR & High accuracy for stereoscopic video & Local spatiotemporal features & PLCC 0.948 and 0.950. \\
\cite{ml-7}\textsuperscript{L} & 2022 & CNN (DenseNet-121) & No-reference; VMAF pre-training & Learns blur and blocking artifacts & PCC up to 0.968; costly for real time. \\
\cite{ml-10}\textsuperscript{L} & 2013 & RBFN & Interpretable; best for small datasets & Non-linear feature--\QoE{} mapping & High accuracy. \\
\cite{ml-20}\textsuperscript{L} & 2012 & Content clustering & Simple; scalable & Clustering of stereoscopic content & 95.4\% success on test sequences. \\
\cite{ml-9}\textsuperscript{L} & 2019 & DNN & Captures complex patterns & Maps network parameters to \QoE{} & Beats prior methods on 80,000+ records. \\
\cite{ml-3}\textsuperscript{L} & 2020 & RFR & Robust ensemble; edge-ready & RFR for streaming \QoE{} & 87.6\% (79\% with mobility). \\
\cite{Mdemaya}\textsuperscript{T} & 2025 & RF & Robust ensemble; edge scheduling & RF for \QoE{}-aware scheduling & Lower latency, higher throughput. \\
\cite{d1}\textsuperscript{L} & 2019 & RFR & Handles complex relationships & Radio metrics (RSSI, RSRP, RSRQ, CINR) & MOS: 94\% accuracy, $R^2$ 0.85. \\
\cite{up-1}\textsuperscript{L} & 2022 & SVM & Facial-expression features & ACR-scale \QoE{} from facial cues & Mean accuracy 0.78. \\
\cite{ml-1}\textsuperscript{L} & 2018 & SVR & Works well with small datasets & Nine frame-level features to VMAF & PLCC 0.98 (VMAF), 0.89 (MOS). \\
\cite{ml-4}\textsuperscript{L} & 2019 & Two-step NN & Physiological and cognitive features & Bitrate, latency and loss to VR \QoE{} & PCC 8.9\% higher than a traditional NN. \\
\cite{ml-5}\textsuperscript{L} & 2020 & DT-Based & Simple; interpretable & DT for VR \QoE{} via immersion and attention & Uses subjective and objective inputs. \\
\cite{up-2}\textsuperscript{L} & 2023 & CL & Adapts to drifting conditions & Continual learning & Regression error 0.9314\%; RMSE 21--86$\times$ lower. \\
\cite{sr-20}\textsuperscript{L} & 2019 & BT & High accuracy on hourly KPIs & Cell-level \QoE{} from LTE KPIs & Accurate throughput and handover estimates. \\
\cite{sr-20}\textsuperscript{L} & 2019 & PROPHET & Trend and seasonality & Forecasts \QoE{} from LTE patterns & Captures load and usage dynamics. \\
\cite{tskan2026}\textsuperscript{L} & 2026 & TSKAN & Inherently interpretable & KANs over DFT features & RMSE 0.2169; beats EBM and NAM. \\
\cite{bilbao2025benchmarking}\textsuperscript{L} & 2026 & Benchmark & Accuracy, efficiency, confidence & SHAP and Conformal Prediction & edRVFL lowest MAE for 10K--100K samples. \\
\cite{osee2025hybrid}\textsuperscript{L} & 2025 & Hybrid Ensemble & Combines QoS and satisfaction & Stacking XGBoost and RF & 90\% F1; SHAP key drivers. \\
\cite{selma2026qoe}\textsuperscript{T} & 2026 & Ridge + HDDE + DDQN & Drift-aware; self-healing & Multimodal fusion and drift detection & $R^2$ 0.953; F1 0.967; $+0.73$ MOS. \\
\cite{zhu2026taolive}\textsuperscript{L} & 2026 & Swin + PWC-Net + DNN & Live-streaming distortions & Optical flow and semantic fusion & PLCC 0.913, SRCC 0.902. \\
\bottomrule
\end{tabular}
\vspace{-0.5em}
\end{table*}

\begin{figure}
    \centering
    \includegraphics[width=0.95\linewidth]{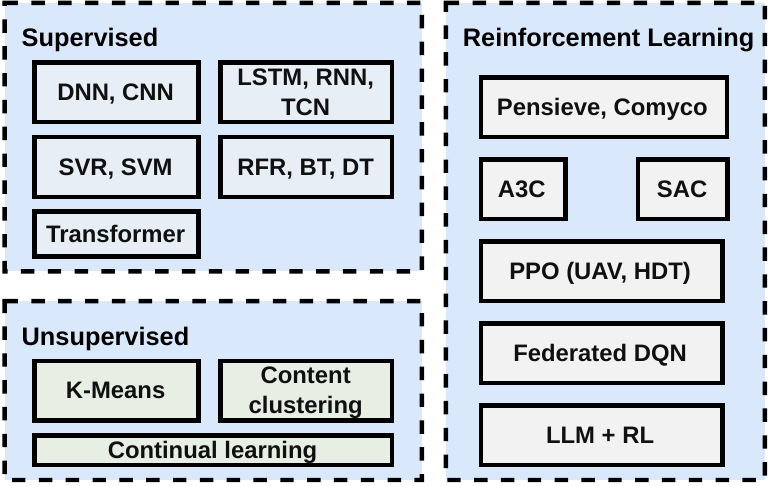}
    \caption{Classification of ML algorithms for \QoE{} management by learning
    paradigm: supervised, unsupervised, and reinforcement learning, all aimed
    at predicting user \QoE{}.}
    \label{fig:classification-ml}
\vspace{-1em}
\end{figure}

This subsection surveys ML algorithms for multimedia \QoE{} management, that is,
applications where the ML model operates on network KPIs, video-streaming
parameters, or user-context data to predict, optimize, or enhance the
user-perceived quality of multimedia services. These applications complement,
rather than duplicate, dedicated perceptual quality-assessment methods such as
large-model video quality assessment~\cite{ge2024lmmvqa}. They span adaptive
bitrate control, immersive-media and cloud-gaming delivery, edge orchestration,
and anomaly detection, and all share the challenge of mapping high-dimensional,
non-stationary states to subjective quality scores.

Each algorithm is analyzed across four deployment-oriented dimensions: data
availability (required KPIs and operational accessibility), generalization
(transferability across networks and content types), interpretability
(explainability to operators), and deployment constraints (inference latency
and compute compatibility with real-time multimedia delivery). Algorithms are
classified into supervised, unsupervised, and reinforcement
learning~\cite{taheri2023machine}. \autoref{fig:classification-ml} categorizes
them, and Tables~\ref{tab:summl} and~\ref{tab:summl-rl} summarize key
approaches. \autoref{sec:ml-supervised} reviews supervised and unsupervised
algorithms, \autoref{sec:ml-rl} reinforcement learning, and
\autoref{sec:ml-gen} generative AI and LLMs. ML is compared with statistical
approaches in \autoref{sec:ml-vs-stat}, as part of model selection.

\begin{figure*}
    \centering
    \includegraphics[width=\linewidth]{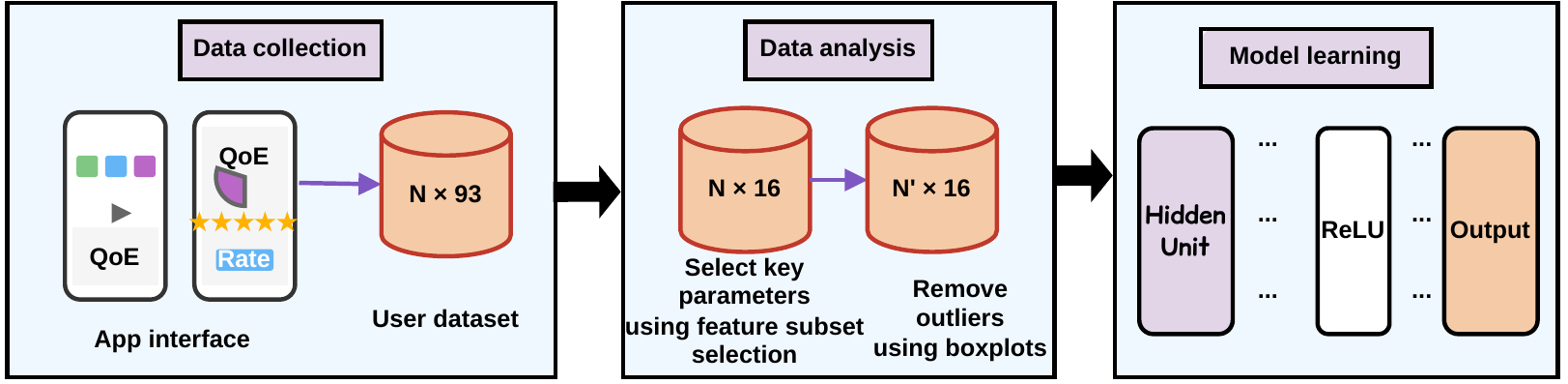}
    \vspace{-1em}
    \caption{DNN-based \QoE{} pipeline: app-based data collection, selection of
    key network features, outlier removal, and model training~\cite{ml-9}.}
    \label{fig:ml-9}
\vspace{-1.5em}
\end{figure*}

\ignore{
\begin{figure*}
    \centering
    \includegraphics[trim=0.2cm 1.5cm 0.2cm 0.2cm, clip, width=1\linewidth]{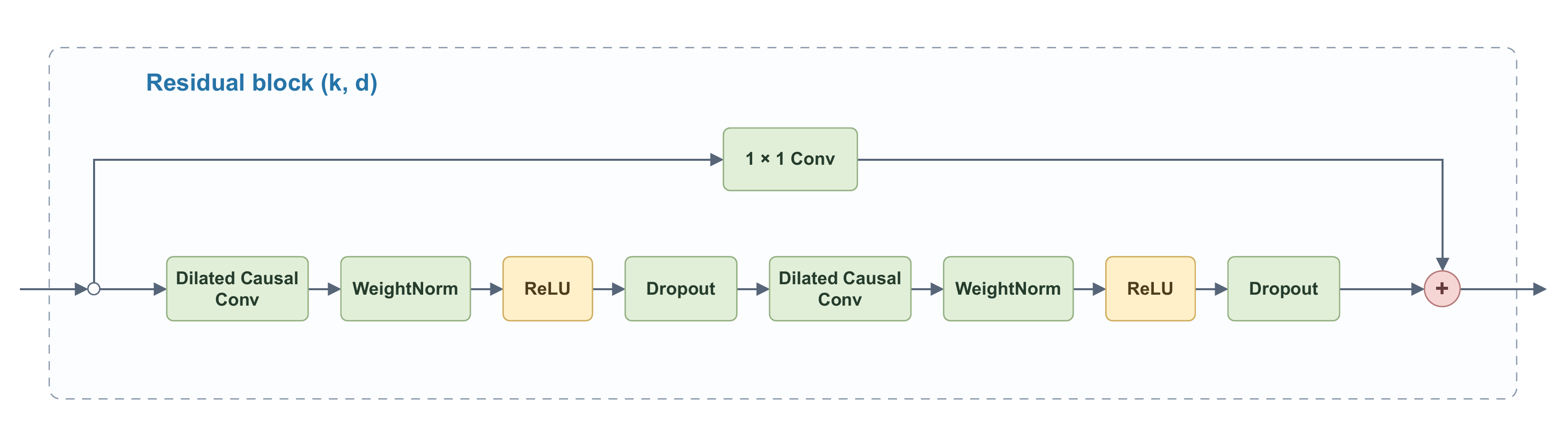}
    \vspace{-2em}
    \caption{\red{Residual block in the  TCN
    architecture~\cite{ml-18}.}}
    \label{fig:ml-18-TCN}
    \vspace{-1em}
\end{figure*}
}

\begin{figure}
    \centering
    \includegraphics[width=\columnwidth]{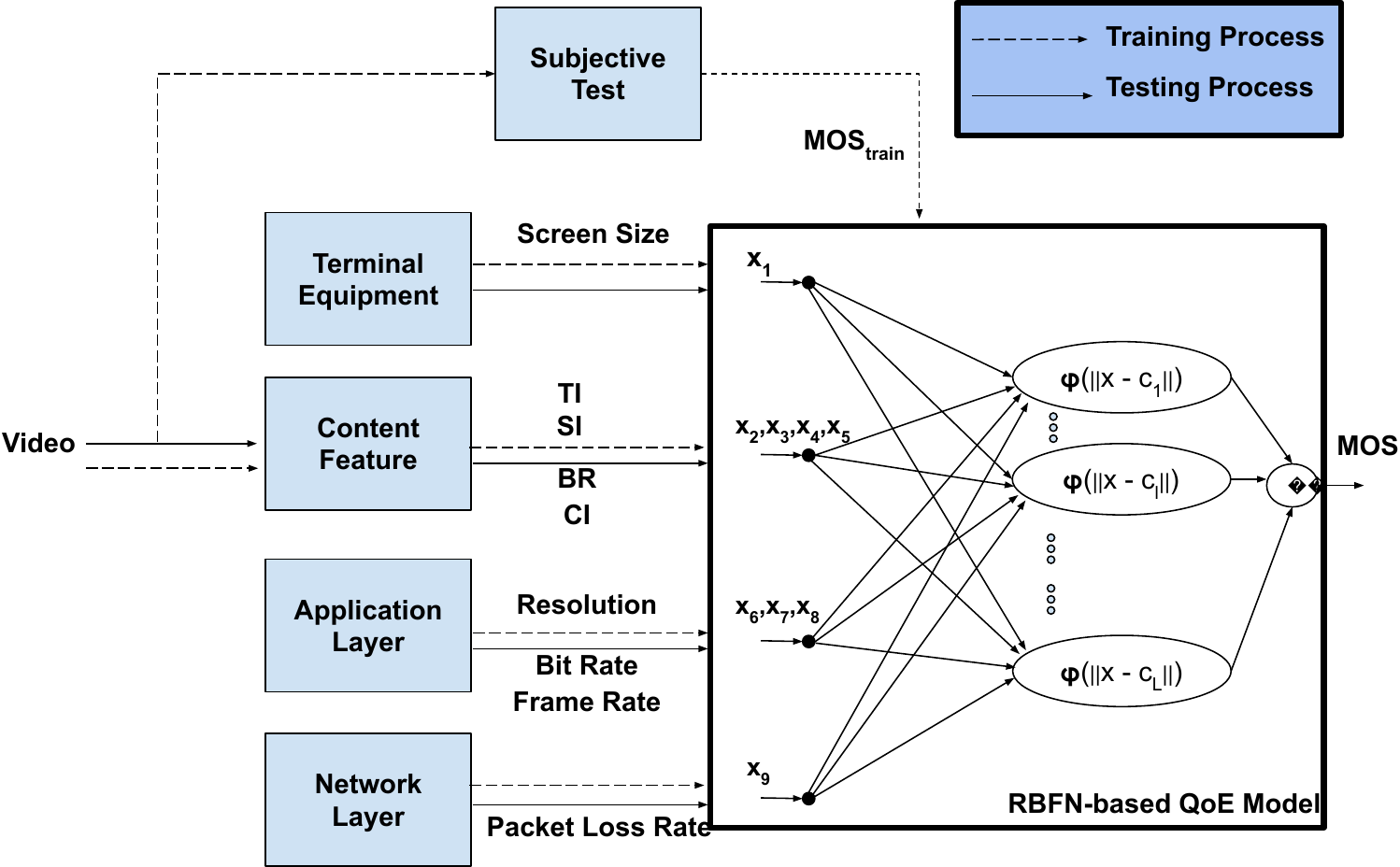}
    \caption{Radial basis function network model for estimating
    video-streaming \QoE{}~\cite{ml-10}.}
    \label{fig:ml-10-RBFN}
\vspace{-1.25em}
\end{figure}

\subsubsection{Supervised and Unsupervised Learning Algorithms}
\label{sec:ml-supervised}\label{sec:ml-unsupervised}

Deep neural networks (DNNs) model the non-linear mapping from objective QoS
metrics (layers 3--4) to subjective \QoE{} (layer 7); Tao \textit{et al.}\ learn this
mapping from more than 80,000 records with 89 network parameters, collected by
a phone app~\cite{ml-9}~(\autoref{fig:ml-9}). High compute cost limits edge
deployment, which favors lighter models in resource-constrained settings.

Convolutional neural networks (CNNs) extract spatial features such as blocking
and blurring that directly affect visual quality~\cite{ml-7,ml-8}. The LSTM
model of~\cite{ml-6} combines quality scores from existing metrics with the
playback state for adaptive streaming, at the cost of substantial labeled
training data.

RNNs with LSTM capture temporal \QoE{} evolution and address the recency effect
in video sessions. LSTMs handle the vanishing gradient problem with memory cells
and reach a linear correlation of up to 0.947 with subjective
scores~\cite{ml-6}; a bidirectional variant reaches 0.894 on LIVE-Netflix, above
the unidirectional LSTM~\cite{ml-21}. Their training cost limits large-scale
real-time deployment.

Temporal convolutional networks (TCNs) offer a more efficient alternative. An
improved TCN infers three times faster than an LSTM (0.67~ms against 2.0~ms on a
PC), although it has more parameters. Its PCC ranges from 0.82 to 0.89 on three
databases~\cite{ml-18}

Transformer architectures use self-attention to capture long-range dependencies
in streaming data. Recent work~\cite{Hamidi} outperforms \texttt{ITU-T P.1203}
with robust cross-device generalization by modeling temporal patterns in
streaming KPIs.

Support vector regression (SVR) generalizes well with limited training data.
With a Gaussian kernel, it maps nine frame-level image features of gaming video
to VMAF, with a PLCC of 0.98 against VMAF and 0.89 against MOS~\cite{ml-1},
though computational cost scales poorly with large datasets.

Support vector machines (SVM) predict user \QoE{} on the ACR scale from facial
expression features extracted during WebRTC conversations, with mean accuracy
0.78~\cite{up-1}.

Random forest regression (RFR) reduces overfitting through ensemble averaging
and achieves high accuracy from radio metrics (RSSI, RSRP, RSRQ, CINR) measured
on smartphones~\cite{d1} and from round-trip time and throughput in emulated 5G
streaming~\cite{ml-3}.

Continual learning (CL) adapts to non-stationary distributions without
catastrophic forgetting, reducing RMSE by up to 86$\times$ relative to several
existing approaches~\cite{up-2}.

Boosted trees (BT) estimate cell-level \QoE{} for the current hour from LTE KPIs
(throughput and handover rates) with selected non-collinear
features~\cite{sr-20}.

PROPHET forecasts future \QoE{} metrics from LTE time series by decomposing
trend, seasonality, and holiday components, which suits regular network dynamics
but is less effective for highly irregular patterns~\cite{sr-20}.

Content clustering groups stereoscopic videos by their spatio-temporal activity
in depth layers, and a separate predictor per cluster reaches a 95.4\% measure
of success on test sequences~\cite{ml-20}. Its simplicity and scalability suit
real-time \QoE{} categorization and resource allocation for large datasets.

\begin{table*}[!tp]
\centering
\caption{{Reinforcement learning, language-model, distributed and other
approaches for \QoE{} prediction and control: core concepts and reported
results. Superscripts give the code of each work
(\autoref{sec:method-coding}).}}
\label{tab:summl-rl}
\renewcommand{\arraystretch}{1.3}
\setlength{\tabcolsep}{4pt}
\footnotesize
\rowcolors{2}{gray!10}{white}
\begin{tabular}{@{}>{\raggedright\arraybackslash}m{0.9cm}
                  >{\centering\arraybackslash}m{0.7cm}
                  >{\raggedright\arraybackslash}m{2cm}
                  >{\raggedright\arraybackslash}m{3.5cm}
                  >{\raggedright\arraybackslash}m{5cm}
                  >{\raggedright\arraybackslash}m{4cm}@{}}
\toprule
\rowcolor{headerblue}
{\color{white}\textbf{Ref.}} & {\color{white}\textbf{Year}} &
{\color{white}\textbf{Algorithm}} & {\color{white}\textbf{Advantage}} &
{\color{white}\textbf{Key Idea}} & {\color{white}\textbf{Results}} \\
\midrule
\cite{ml-19}\textsuperscript{L} & 2020 & GAN & Removes compression artifacts frame by frame & Enhances visual quality of gaming images & Improvement in perceived image quality (up to about 2 MOS). \\
\cite{panahi2025bridging}\textsuperscript{L} & 2025 & LLM + RF & Dual-layer assessment; semantic analysis & Combines objective network KPIs with subjective feedback using LLMs & Scalable subjective scoring from live comments. \\
\cite{wang2025llm4band}\textsuperscript{T} & 2025 & RL + LLM & Enhanced reinforcement learning & Uses LLMs to improve bandwidth estimation in RL agents & +5.5--12.4\% estimation gain; +21\% call quality vs.\ GCC (simulation). \\
\cite{panahi2025anomaly}\textsuperscript{--} & 2026 & Hybrid/LLM & \QoE{}-driven anomaly detection & Survey of paradigms including simulation, measurement, and hybrid approaches & Identifies challenges in real-time implementation and subjective--objective synergy. \\
\cite{rezaee2025realtime}\textsuperscript{T} & 2025 & AI Agents & Real-time monitoring & Uses AI agents to monitor multiuser communications & Concept only; no evaluation reported. \\
\cite{SHA202565}\textsuperscript{--} & 2025 & Enhanced GA & \QoE{}-aware infrastructure optimization & Enhanced GA for \QoE{}-optimal edge-server placement (coverage vs.\ resources) & 7--34\% \QoE{} gain over five baselines (simulation). \\
\cite{zhou2025federated}\textsuperscript{D} & 2026 & Federated Transformer & Privacy-preserving RL for VR streaming & Decision transformer with prompts in a federated setting across MEC servers & Optimizes resource allocation without central data collection; maintains high \QoE{}. \\
\cite{garcia2025decentralized}\textsuperscript{M} & 2027 & Blockchain + FHE & End-to-end privacy; decentralized trust & Smart contracts and Fully Homomorphic Encryption for \QoE{} computation & \QoE{} computed on encrypted data; on-chain confirmation takes 8.8--11.5~s. \\
\cite{congia2026towards}\textsuperscript{L} & 2026 & BERT (Fine-tuned) & Language models as \QoE{} generative models & Text-to-score regression using natural language descriptions & MAE 0.36, RMSE 0.51 on blended dataset; degrades on unseen IFs. \\
\cite{lyu2026drllms}\textsuperscript{T} & 2026 & RL + CoT Control & Optimizes \QoE{} for LLM streaming & Network-adaptive chain-of-thought length selection via grid search & Trades answer quality for responsiveness (TTFA) as bandwidth varies. \\
\cite{karapantelakis2026qoe}\textsuperscript{T} & 2026 & LLM-based Agents & \QoE{} assurance for encrypted 5G traffic & Deployment Agent maps \QoE{} to QoS; Monitoring Agent enforces rules & 96\% accuracy in selecting the correct control action (AR testbed). \\
\cite{perez2026video}\textsuperscript{--} & 2026 & Position Paper & Adapts video \QoE{} model to LLM services & Identifies shift from content to interaction, compute, and multidimensional outcomes & Proposes a research agenda for LLM \QoE{}. \\
\cite{yang2025joint}\textsuperscript{T} & 2025 & A3C & Joint layering and compute allocation & Tile-based SVC; cloud--edge--end; dual-connectivity & Higher \QoE{} and lower latency than AC and fixed baselines. \\
\cite{wen2026immersive}\textsuperscript{D} & 2026 & SAC & Per-frame radio--compute--content control & Near-RT RIC xApp; content-hit ratio $\phi$; Weber--Fechner \QoE{} & $>11\%$ median MTP reduction vs.\ PF-AO (simulation); higher fairness. \\
\cite{zhang2025realtime}\textsuperscript{D} & 2025 & Transformer + Federated DQN & Multi-view sports; federated scheduling & Viewpoint prediction; \QoE{}-aware stream ranking; FedAvg & Top-1 84.6\%; $-24\%$ latency; $+10\%$ \QoE{} (different baselines). \\
\cite{chen2026continual}\textsuperscript{D} & 2026 & PPO + GNN + continual & UAV-assisted 360° VR; handles GU churn & NOMA + SVC; dynamic UAV-GU graph; stable/unstable node decomposition & 6--12\% reward gain; 40--60\% traffic reduction. \\
\cite{wu2025qoe}\textsuperscript{T} & 2025 & PPO (VisHap) & Joint visual + haptic; HDT & Adaptive JND; multi-modal \QoE{} model & \QoE{} 0.772 vs 0.492--0.705 for baselines. \\
\bottomrule
\end{tabular}
\vspace{-1em}
\end{table*}

\subsubsection{Reinforcement Learning Algorithms}
\label{sec:ml-rl}

Reinforcement learning (RL) algorithms learn optimal policies through
interaction with the environment. This makes them well suited to direct \QoE{}
optimization when the reward signal is defined in terms of user-perceived
quality. Pensieve~\cite{mao2017pensieve} established RL as a viable approach for
\QoE{}-driven adaptive bitrate selection and outperforms classical heuristics by
directly maximizing a \QoE{} reward that combines bitrate, rebuffering, and
quality smoothness. Comyco~\cite{huang2019comyco} improves sample efficiency
through imitation learning; both run as a service on an ABR server during
playback. Oboe, a non-learned counterpart, is discussed with loose coupling in
\autoref{sec:loose}. Recent work~\cite{wang2025llm4band} combines RL with LLMs
for bandwidth estimation in real-time communication, where better estimates
raise call quality.

Six recent RL controllers act on radio, compute and content resources rather
than on the bitrate alone: A3C for layering and power~\cite{yang2025joint}, SAC
in a RIC xApp~\cite{wen2026immersive}, PPO on UAVs~\cite{chen2026continual}, PPO
for visual-haptic delivery~\cite{wu2025qoe}, a federated decision
transformer~\cite{zhou2025federated} and a federated multi-agent
DQN~\cite{zhang2025realtime}. Their formulations are described with their
coupling pattern in \autoref{sec:tight} and \autoref{sec:distributed}, because
where each controller runs matters more for \QoE{} than the RL algorithm it
uses.

\subsubsection{Generative AI and LLMs}
\label{sec:ml-gen}

Emerging research explores generative AI and LLMs for \QoE{} assessment. Panahi
\textit{et al.}~\cite{panahi2025bridging} introduced a dual-layer framework that
integrates objective network modeling with subjective LLM-based comment
analysis. LLM4Band~\cite{wang2025llm4band} enhances RL with LLMs for accurate
bandwidth estimation. In simulated real-time calls, it reports estimation gains
of 5.5\% to 12.4\% over the best baselines in three scenarios and a 21\% gain in
call quality over the GCC heuristic at low bandwidth. Beyond using LLMs merely
as analysis tools, recent works explore their role as generative models and
their application in new service paradigms. Congia \textit{et al.}~\cite{congia2026towards}
investigate whether language models can act as \QoE{} generative models by
predicting MOS from textual descriptions of application scenarios. Using a
fine-tuned BERT model on three heterogeneous video streaming datasets, they
achieve an MAE of 0.36 and RMSE of 0.51 on a blended dataset, demonstrating that
language models can approximate subjective \QoE{} with promising reliability,
though their performance degrades on unseen influence factors.

Generative adversarial networks (GANs) enhance the visual quality of compressed
gaming video: a modified SRGAN removes compression artifacts frame by frame and
raises perceived image quality in a subjective test~\cite{ml-19}. Their compute
cost and training instability make real-time deployment challenging.

\subsection{\underline{Summary and Lessons Learned}}
\label{sec:landscape-summary}

Tables~\ref{tab:closed-source}--\ref{tab:summl-rl} mark each work with its code.
Three lessons follow. First, measurement and learning are mostly separate
artifacts: 15 of the 21 frameworks in Tables~\ref{tab:closed-source}
and~\ref{tab:open-source} train no model. Their outputs are stored logs,
spreadsheets, or databases that suit an offline model more than a live one.
Second, most ML models in Tables~\ref{tab:summl} and~\ref{tab:summl-rl} are
trained and tested offline on datasets. They report results with different
metrics, so a single accuracy ranking is not possible. Third, the models that
act on the service at run time are recent RL and agent-based controllers, half
of them evaluated only in simulation. How the two components are joined is the
subject of \autoref{sec:patterns}.

\section{Framework--ML Integration Patterns}
\label{sec:patterns}

\begin{figure*}[]
    \centering
    \includegraphics[width=\linewidth, clip,trim={0cm 0.1cm 0cm 0cm}]{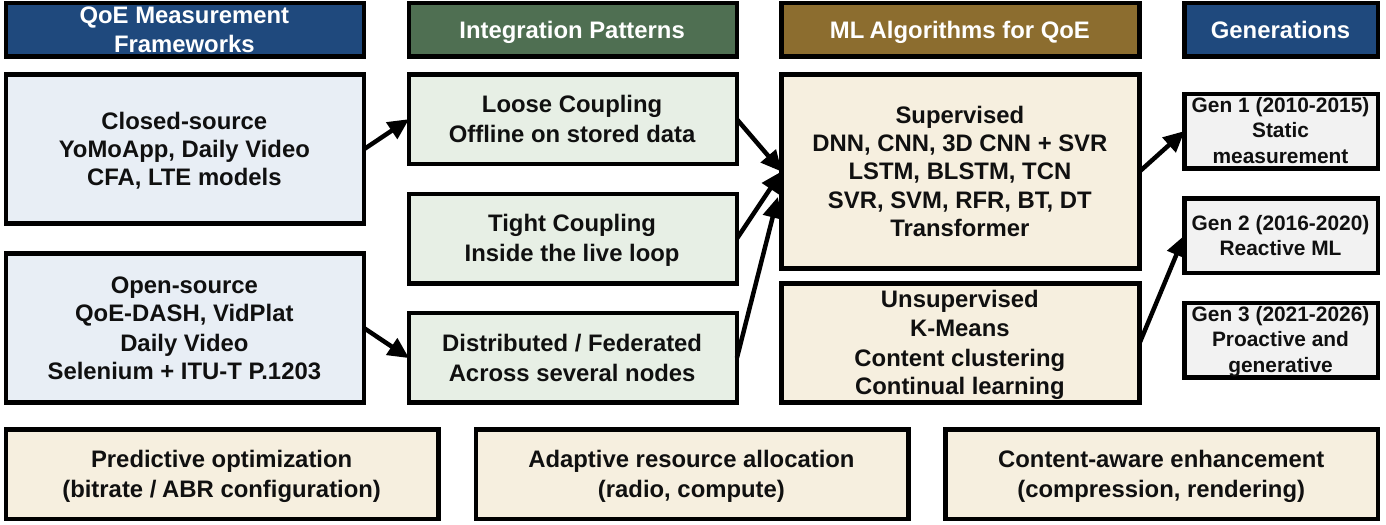}
    \vspace{-1em}

    \caption{Framework-ML integration taxonomy. \QoE{} measurement frameworks,
    both closed- and open-source, connect to ML algorithms through three
    integration patterns (loose coupling, tight coupling, and
    distributed/federated), which drive three technical enhancement mechanisms.
    The right-side timeline traces the generational evolution from static
    measurement (2010--2015) through reactive ML (2016--2020) to proactive and
    generative solutions (2021--2026).}
    \label{fig:framework-overview}
\vspace{-1em}
\end{figure*}

This section answers \texttt{RQ1}: \emph{how are \QoE{} measurement frameworks coupled with ML models?}%
It covers the link between stages~2 and~3 of \autoref{fig:all-in-one},
which \autoref{sec:landscape} left open. Each pattern is defined by where model
inference runs relative to the measurement loop. The subsections follow that
order: offline on stored data (\autoref{sec:loose}), inside the live loop
(\autoref{sec:tight}) and across several nodes (\autoref{sec:distributed}). Of
the 68 reviewed works, 31 are loose, 12 tight and 4 distributed; 17 are
measurement-only works that support a pattern without instantiating it. Four
receive no pattern code. \autoref{fig:framework-overview} summarizes the
taxonomy. \autoref{sec:patterns-summary} compares the patterns.

\subsection{\underline{Loose Coupling}}
\label{sec:loose}

\emph{Loose coupling (framework as data collector).} In this pattern, frameworks
serve as data collectors. Tools such as YoMoApp~\cite{sr-1,sr-12} and Daily
Video~\cite{sr-18} record player states and user ratings and store them for
later analysis. They train no model themselves; their stored data can feed
separate processing pipelines for feature extraction and offline ML training.
This approach characterizes early \QoE{} studies (Gen 1 and Gen 2), where the
priority was establishing ground-truth datasets for model development rather
than real-time adaptation. The disconnect between measurement and inference
introduces latency that precludes sub-second adaptation.

Loose coupling is the most common pattern by a wide margin. It covers almost
every prediction model in Tables~\ref{tab:summl} and~\ref{tab:summl-rl}, which
are trained and tested on datasets (\autoref{sec:ml-prediction}). It also covers
the framework works that train a model on their own collected data: the neural
network of~\cite{sr-11}, the decision tree on smartphone measurements
of~\cite{sr-17}, the LTE models of~\cite{sr-20} and the network-only MOS model
of~\cite{mine-archive}. Of the 31 loose works, 19 use laboratory data, 7 field
data, 2 simulation and one a production service; 2 do not state their data
source.

In 5G O-RAN, Carballo Gonz{\'a}lez \textit{et al.}~\cite{carballo2025oran} train a
Gradient Boosting model that estimates session \QoE{} from network-side data
alone, so that the operator can later allocate resources by \QoE{} without
asking the service provider for a \QoE{} estimate at run time; the model reaches
$R^2 = 0.986$ and RMSE $= 0.091$ on link-level simulation data. The service
provider still labels the training data. Running the model as a RIC xApp is
proposed but not tested.

A deployed instance sharpens the loose-coupling pattern. LingXi~\cite{jia2025lingxi}
runs on Kuaishou as an asynchronous plugin on top of the production ABR
algorithm: it consumes player telemetry (bitrate, throughput, stall history,
stall-exit history), predicts the segment-level exit rate, and adjusts the
\emph{objective parameters} of the underlying ABR algorithm rather than the
bitrate decision itself; the parameter tuning runs in a low-priority background
thread. The measurement loop and the inference loop are therefore decoupled by
design, and the ABR algorithm keeps its own stability mechanisms intact. This
survey names this arrangement \emph{objective tuning} and records it as an
emerging direction (\autoref{sec:emerging}), because LingXi is its only learned
instance. It is supported by production-scale evidence: a 10-day
difference-in-differences A/B test over $8\%$ of Kuaishou traffic reports a
statistically significant $+0.146\%$ increase in total watch time, a $+0.103\%$
increase in bitrate, and a $-1.287\%$ decrease in stall time, with up to a
$-15\%$ decrease in stall time for users in bandwidth-constrained conditions.

Oboe~\cite{akhtar2018oboe} addresses generalizability by auto-tuning to
prevailing network conditions without a learned model: it computes offline the
best parameters of an existing ABR algorithm for each network state, then
detects state changes during playback and applies the stored parameters. It
therefore tunes the objective rather than the bitrate, as LingXi does. It was
tested on a testbed and in a two-week pilot.

\subsection{\underline{Tight Coupling}}
\label{sec:tight}

\emph{Tight coupling (embedded intelligence).} This pattern embeds ML inference
engines directly within the measurement loop. The CFA algorithm~\cite{sr-3}
exemplifies this approach: a backend relearns critical session features every 30
minutes. Front-end servers of the content provider answer each arriving client
in under a millisecond. The LTE framework of~\cite{sr-20}, often grouped here,
trains and tests its models offline and is therefore coded loose. Processing
data locally achieves ultra-low latency and enables immediate reaction to
transient network fluctuations. This integration suits latency-sensitive
applications such as cloud gaming, although none of the reviewed cloud gaming
works yet runs a model in the loop (\autoref{sec:domains}).

Twelve works run a model inside the live loop. Two choose the ABR bitrate on a
server during playback. Pensieve is trained by RL in a simulator. Comyco is
trained by imitation learning~\cite{mao2017pensieve,huang2019comyco}. Others
decide placement or policy at run time: a Random-Forest Kubernetes
scheduler~\cite{Mdemaya}, the LLM-based bandwidth estimator of LLM4Band inside
the call loop~\cite{wang2025llm4band}, the reasoning-length optimizer of
DRLLMS~\cite{lyu2026drllms}, the LLM agents of~\cite{karapantelakis2026qoe} and
the self-healing controller of QoE-Foresight~\cite{selma2026qoe}
(\autoref{sec:challenges-key}). The Sandvine product~\cite{sr-4} claims ML
inside a live analytics pipeline without reporting an evaluation. The AI agents
of~\cite{rezaee2025realtime} are a concept without evaluation. The evidence is
thin: one production pilot (CFA), five laboratory or testbed studies, four
simulations and two works without an evaluation. Two RL formulations show the
range of the pattern.

Beyond ABR selection, RL has been applied to the joint control of layered
encoding and compute allocation for VR delivery. Yang \textit{et al.}~\cite{yang2025joint}
formulate tile-based scalable VR video transmission in a 6G cloud--edge--end
architecture, cache the tile sets at MEC nodes, and use an Asynchronous
Advantage Actor-Critic (A3C) agent to jointly allocate the sub-6~GHz and
millimeter-wave link resources, the edge decoding power, and the terminal
decoding power under an end-to-end latency constraint. The reported simulations
cover 8 to 16 VR users over 6 MEC servers, with tile-based scalable encoding in
a $4\times8$ grid, and the authors report higher average \QoE{} (measured as
PSNR) and lower latency than an Actor-Critic baseline, a fixed-allocation
baseline, and configurations without viewpoint prediction. The same paper also
reports that the A3C agent attains higher average \QoE{} and lower latency as
MEC computation power and UE computation power increase.

PPO, also used by the UAV controller of~\cite{chen2026continual}
(\autoref{sec:distributed}), has been applied outside the video-only case. Wu et
al.~\cite{wu2025qoe} address joint visual and haptic feedback in a human digital
twin (HDT), and design a \QoE{} model that combines video quality, continuous
video quality switching rate, and average haptic feedback error. The controller,
VisHap, sets the bandwidth allocation, the bitrate and rendering mode of each
video tile, and the haptic signal's compression threshold (JND) on every slot,
and it is trained with PPO under a synchronization constraint between the visual
and haptic streams. On a dataset collected from a HDT multi-modal interaction
platform, VisHap reaches an average \QoE{} of $0.772$, against $0.492$--$0.705$
for the five evaluated baselines. The authors also report that adaptive JND
adjustment outperforms fixed JND values ($13\%$, $18\%$, $23\%$, $28\%$) and that
the visual and haptic delays remain within the synchronization bounds on every
slot shown.

\subsection{\underline{Distributed and Federated Integration}}
\label{sec:distributed}

\emph{Distributed/federated integration.} Platforms such as \QoE{}-DASH~\cite{sr-21}
and VidPlat~\cite{sr-6} emulate many users or recruit many crowd workers, which
yields diverse datasets, but neither trains a model; they are measurement tools
that can support this pattern. Federated learning trains local models on the
device and shares only gradients with a central server. This supports general
data protection regulation (GDPR) compliance and improves generalizability
because the model is trained on heterogeneous real-world conditions rather than
on lab environments. Recent works apply this pattern to complex multimedia
scenarios. Zhou \textit{et al.}~\cite{zhou2025federated} propose FedPromptDT, a federated
prompt-based decision transformer for resource allocation in customized VR
streaming. By combining federated learning with sequence modeling, they
pre-train a generalized policy across multiple MEC servers to optimize \QoE{}
(considering latency, user attention, and preferred resolutions) without
centralizing sensitive user viewing data. Furthermore, Garcia et
al.~\cite{garcia2025decentralized} push the boundary of privacy by proposing a
decentralized \QoE{} system for 6G networks that combines blockchain with fully
homomorphic encryption (FHE). This allows service providers to compute aggregate
\QoE{} metrics directly on encrypted user data via smart contracts, with
end-to-end privacy as the design goal; automated compliance verification is left
to future work. The system computes \QoE{} without a learned model, so this
survey codes it as measurement only: it supports the pattern rather than
instantiating it.

Only four reviewed works distribute training or decisions over several nodes.
All four were published in 2025 or 2026. All are evaluated in simulation, one of
them also on a testbed: FedPromptDT~\cite{zhou2025federated}, the RIC-based
controller of~\cite{wen2026immersive}, the federated scheduler
of~\cite{zhang2025realtime} and the UAV controller of~\cite{chen2026continual}.
Each takes a different route, described below.

An emerging direction of the distributed pattern organizes the \QoE{} loop
around the RAN itself. The O-RAN-assisted volumetric playback framework of Wen
\textit{et al.}~\cite{wen2026immersive} places the joint radio--compute--content
controller inside a Near-RT RIC xApp, so that bandwidth, transmit power, and
O-Cloud compute are decided within the same control loop as the measurement and
rendering pipeline. This arrangement does not fit the loose-coupling or
tight-coupling patterns cleanly, because the controller is neither a separate
offline data collector nor an inference engine embedded in a single measurement
probe; it is an operator-side control plane that consumes telemetry through E2
and acts on the RAN through the same interface, while O-Cloud compute is
requested through the service management and orchestration (SMO) layer. The
direction is named here \emph{RIC-mediated integration}; Wen \textit{et al.}\ are its
only tested instance and the O-RAN model of~\cite{carballo2025oran} proposes the
same placement without testing it.

A different RL formulation targets the radio--compute--content loop at the RAN.
Wen \textit{et al.}~\cite{wen2026immersive} propose an O-RAN-integrated playback framework
for immersive volumetric video in which a Near-RT RIC xApp coordinates the
O-Cloud compute resources, the gNB transmit power, and the radio bandwidth on a
per-frame basis. The content-hit ratio $\phi_u$ is treated as a continuous
control variable under a Weber--Fechner \QoE{} model, and a soft actor-critic
(SAC) agent with structured action decomposition and \QoE{}-aware reward shaping
solves the resulting high-dimensional control problem. The framework is
implemented on a 5G O-RAN testbed built from OpenAirInterface, FlexRIC, an Ettus
USRP X300, and a Meta Quest 3 headset, and the authors report, in simulation,
that SAC reduces median motion-to-photon latency by more than $11\%$ relative to
the PF-AO baseline while improving fairness; the testbed measurements corroborate
the latency-throughput trade-offs.

Federated orchestration has also been applied at the level of the streaming
scheduler rather than the \QoE{} model. Zhang~\cite{zhang2025realtime} proposes a
collaborative edge framework for multi-view sports streaming in which a
transformer-based viewpoint predictor, a \QoE{}-aware stream selector, and a
federated multi-agent DQN scheduler are coordinated across edge nodes via
FedAvg. The author reports a Top-1 viewpoint accuracy of $84.6\%$, a $24\%$
latency reduction, and a $10\%$ \QoE{} improvement, each measured against a
different baseline, and a \QoE{} drop of $7.9\%$ when the large model is replaced
by a GRU. This direction, named here \emph{federated scheduling}, extends the
distributed/federated integration pattern from privacy-preserving \QoE{}
inference to federated content-selection and scheduling.

A third RL setting is UAV-assisted 360° VR delivery, where the radio link and
the content pipeline both change on a per-slot basis. Chen et
al.~\cite{chen2026continual} consider UAVs that capture, encode, and stream
multi-quality tiled 360° VR video to ground users, and they combine
non-orthogonal multiple access (NOMA) with scalable video coding (SVC) so that
multiple users are served from the same downlink transmission with different
quality levels. A FoV-aware PPO agent chooses the user scheduling and the video
quality levels, and an inner monotonic-optimization layer allocates the transmit
power for the chosen discrete actions. Because users join and leave the network,
the authors further propose DGC-DRL, which models the evolving UAV-GU network as
a dynamic heterogeneous graph and adds a continual-learning module with
stable/unstable node decomposition, parameter isolation, and replay. The authors
report a 6--12\% reward gain over the evaluated baselines, and a 40--60\%
reduction in aggregate traffic when the tiled SVC variant is used.

\subsection{\underline{Summary and Lessons Learned}}
\label{sec:patterns-summary}

\begin{table*}[]
\centering
\caption{{Comparison of the three coupling patterns over the 68 reviewed works. Evidence counts the strongest evaluation reported by each work.}}
\label{tab:patterns}
\renewcommand{\arraystretch}{1.3}
\setlength{\tabcolsep}{4pt}
\footnotesize
\begin{tabular}{@{}>{\raggedright\arraybackslash}m{1.5cm}
                  >{\raggedright\arraybackslash}m{3.5cm}
                  >{\centering\arraybackslash}m{0.75cm}
                  >{\raggedright\arraybackslash}m{5cm}
                  >{\raggedright\arraybackslash}m{2.8cm}
                  >{\raggedright\arraybackslash}m{3.2cm}@{}}
\toprule
\rowcolor{headerblue}
{\color{white}\textbf{Pattern}} & {\color{white}\textbf{Where inference runs}} & {\color{white}\textbf{Works}} &
{\color{white}\textbf{Evidence}} & {\color{white}\textbf{Strength}} & {\color{white}\textbf{Limitation}} \\
\midrule
Loose & Offline, on stored data & 31 & 19 laboratory, 7 field, 2 simulation, 1 production, 2 not stated & Large datasets; no run-time cost & No reaction within a session \\
Tight & Inside the live loop, on the client, a server, or the network & 12 & 1 production pilot, 5 laboratory or testbed, 4 simulation, 2 none & Reacts within a session & Little field evidence; run-time cost \\
Distributed & Across several nodes that share updates or decisions & 4 & 4 simulation (1 also testbed) & Privacy; scale across nodes & No field evidence yet \\
\bottomrule
\end{tabular}
\vspace{-0.75em}
\end{table*}

\autoref{tab:patterns} compares the three patterns. Three lessons answer \texttt{RQ1}.
First, integration in the reviewed literature is mostly loose: 31 of the 47
works that use ML train and test their models offline. Second, the closer a
pattern moves to the live loop, the weaker its evidence: tight coupling has one
production pilot. Distributed integration has none. Third, the two production
results come from opposite patterns, CFA (tight) and LingXi (loose), so
production readiness does not follow from the pattern alone; both deployments
sit with the content provider, which owns both the measurement and the control
points. Which \QoE{} actions each pattern supports is the subject of
\autoref{sec:mechanisms}.

\section{Enhancement Mechanisms}
\label{sec:mechanisms}

This section answers \texttt{RQ2}: \emph{which \QoE{} actions does each coupling
pattern support in the reviewed works?} It covers stages~4 and~5 of
\autoref{fig:all-in-one}, the optimizer and the service adaptation, which turn a
prediction into an action. The reviewed works act on the service through three
mechanisms, ordered from the session outward: the bitrate and configuration of a
session, the resources of the network, and the content itself.

Three mechanisms are identified through which integrated solutions enhance
\QoE{}. \autoref{tab:mechanisms} maps each mechanism to the reviewed works that
act on the service, with their coupling pattern and evidence level.

\emph{Predictive optimization.} Sequence models (LSTM~\cite{ml-6},
TCN~\cite{ml-18}) predict \QoE{} continuously over a session, which is the input
that a proactive adaptive bitrate (ABR) controller needs. Both are evaluated
offline; the TCN authors name a client-side switching controller only as future
use. The reviewed works that close this loop act on the bitrate or on the ABR
configuration: Pensieve and Comyco choose the bitrate in the loop, LingXi and
Oboe tune the ABR objective, QoE-Foresight issues bitrate actions in trace
replay, LLM4Band feeds rate control in calls and DRLLMS sets the reasoning
length of an LLM service. This shifts \QoE{} management from reactive to
proactive.

\emph{Adaptive resource allocation.} Prediction models such as RFR~\cite{ml-3}
and the LTE models of~\cite{sr-20} estimate \QoE{} from network KPIs, which is
the input that allocation needs; neither performs the allocation itself. The IoT
framework of~\cite{sr-19} supplies MOS ratings that can serve as allocation
targets. Edge-server placement by a genetic algorithm~\cite{SHA202565} raises
simulated \QoE{} by 7--34\% as an offline planning step. A Random-Forest
scheduler~\cite{Mdemaya} lowers latency and raises throughput on a small
Kubernetes cluster. Run-time allocation by a learned controller appears in
immersive delivery: A3C for layering and power~\cite{yang2025joint}, SAC in a
RIC xApp~\cite{wen2026immersive}, a federated decision
transformer~\cite{zhou2025federated}, PPO on UAVs~\cite{chen2026continual}, PPO
for visual-haptic bandwidth~\cite{wu2025qoe} and LLM agents that enforce QoS
rules~\cite{karapantelakis2026qoe}. The O-RAN model of~\cite{carballo2025oran}
(\autoref{sec:loose}) prepares this step but does not perform it.

\emph{Content-aware enhancement.} A GAN~\cite{ml-19} removes compression
artifacts from gaming frames and raises perceived quality without more
bandwidth; it is evaluated offline, frame by frame. Content-side control also
appears inside run-time controllers: tile layering and scalable coding
levels~\cite{yang2025joint,chen2026continual}, the content-hit
ratio~\cite{wen2026immersive}, the rendering mode and haptic
compression~\cite{wu2025qoe} and stream selection~\cite{zhang2025realtime}.
These controls decouple perceived visual quality from available bandwidth.

\begin{table*}[]
\centering
\caption{{Works that act on the service, by mechanism, with their code (M, L,
T, D, or -- for no learned model) and evidence.}}
\label{tab:mechanisms}
\renewcommand{\arraystretch}{1.3}
\setlength{\tabcolsep}{4pt}
\footnotesize
\begin{tabular}{@{}>{\raggedright\arraybackslash}m{3.2cm}
                  >{\raggedright\arraybackslash}m{7.4cm}
                  >{\raggedright\arraybackslash}m{5.4cm}@{}}
\toprule
\rowcolor{headerblue}
{\color{white}\textbf{Mechanism}} & {\color{white}\textbf{Works (code)}} & {\color{white}\textbf{Evidence}} \\
\midrule
Predictive optimization (bitrate and ABR configuration) & CFA~\cite{sr-3} (T), Pensieve~\cite{mao2017pensieve} (T), Comyco~\cite{huang2019comyco} (T), QoE-Foresight~\cite{selma2026qoe} (T), LLM4Band~\cite{wang2025llm4band} (T), DRLLMS~\cite{lyu2026drllms} (T), LingXi~\cite{jia2025lingxi} (L), Oboe~\cite{akhtar2018oboe} (--) & Production: CFA, LingXi. Laboratory: Pensieve, Comyco, DRLLMS, Oboe. Simulation: QoE-Foresight, LLM4Band. \\
Adaptive resource allocation & Yang~\cite{yang2025joint} (T), VisHap~\cite{wu2025qoe} (T), QoE agents~\cite{karapantelakis2026qoe} (T), HERCULE~\cite{Mdemaya} (T), Wen~\cite{wen2026immersive} (D), FedPromptDT~\cite{zhou2025federated} (D), DGC-DRL~\cite{chen2026continual} (D), edge placement~\cite{SHA202565} (--), IoT estimator~\cite{sr-15} (M), IPTV manager~\cite{sr-8} (M) & Testbed or cluster: Wen, QoE agents, HERCULE, IPTV manager. Simulation: the other six. \\
Content-aware enhancement & GAN enhancer~\cite{ml-19} (L), multi-view stream selection~\cite{zhang2025realtime} (D); content controls inside~\cite{yang2025joint,chen2026continual,wen2026immersive,wu2025qoe} & Laboratory (offline): GAN. Simulation: the others. \\
\bottomrule
\end{tabular}
\vspace{-0.75em}
\end{table*}

\emph{Summary and lessons learned.} \autoref{tab:mechanisms} maps the mechanisms
to the patterns. Predictive optimization of the bitrate or the ABR
configuration is the only mechanism with production evidence. It works under
both loose and tight coupling. Adaptive resource allocation is where the tight
and distributed controllers concentrate, most of them evaluated in simulation.
Content-aware enhancement has one offline enhancer and otherwise appears as one
control among others inside allocation controllers. The demands of these
mechanisms on latency and data differ by service, which is the subject of
\autoref{sec:domains}.

\section{Domain Fit and Model Selection}
\label{sec:domains}

This section answers \texttt{RQ3}: \emph{which coupling patterns have been applied in each
service domain and which service constraints explain these choices?} It covers
stage~1 of \autoref{fig:all-in-one}, the service environment.
\autoref{sec:domain-constraints} relates the constraints of each domain to the
coupling found there. \autoref{sec:synthesis-decision} turns the verified results
into a guide for model selection. \autoref{tab:domains} gives the distribution
of codes by domain.

\begin{table}[]
\centering
\caption{{Reviewed works by service domain and code (68 works).}}
\label{tab:domains}
\renewcommand{\arraystretch}{1.3}
\setlength{\tabcolsep}{3pt}
\footnotesize
\begin{tabular}{@{}>{\raggedright\arraybackslash}m{3.6cm}cccccc@{}}
\toprule
\rowcolor{headerblue}
{\color{white}\textbf{Domain}} & {\color{white}\textbf{Works}} & {\color{white}\textbf{M}} & {\color{white}\textbf{L}} & {\color{white}\textbf{T}} & {\color{white}\textbf{D}} & {\color{white}\textbf{--}} \\
\midrule
Video streaming & 31 & 10 & 15 & 5 & 0 & 1 \\
Live streaming & 2 & 0 & 1 & 0 & 1 & 0 \\
Immersive and 3D media & 11 & 0 & 5 & 3 & 3 & 0 \\
Gaming video and cloud gaming & 5 & 1 & 4 & 0 & 0 & 0 \\
Real-time communication & 3 & 0 & 1 & 2 & 0 & 0 \\
IoT and edge & 3 & 2 & 0 & 1 & 0 & 0 \\
LLM services & 2 & 0 & 0 & 1 & 0 & 1 \\
Mobile networks (service-agnostic) & 11 & 4 & 5 & 0 & 0 & 2 \\
\midrule
\textbf{Total} & \textbf{68} & \textbf{17} & \textbf{31} & \textbf{12} & \textbf{4} & \textbf{4} \\
\bottomrule
\end{tabular}
\vspace{-0.75em}
\end{table}

\subsection{\underline{Domain Constraints and Coupling Choices}}
\label{sec:domain-constraints}

\emph{Multimedia streaming.} The primary limitation is the trade-off between
rebuffering events and bitrate switching frequency. Stalling has the most severe
impact on user satisfaction~\cite{sr-1,sr-12}, and frequent quality fluctuations
also degrade \QoE{}. Current frameworks often struggle to optimize this balance
in highly dynamic mobile networks where bandwidth varies rapidly. Video
streaming holds 31 of the 68 works and both production
results~\cite{sr-3,jia2025lingxi}. Its playback buffer gives a controller
seconds of slack, so the model can run outside the player: on the provider's
servers in CFA and Pensieve, or as an asynchronous plugin in LingXi. Most
streaming works (15 of 31) are nevertheless loose prediction models that stop
before control.

\emph{Live streaming.} Live broadcasts add constraints that video on demand
lacks: frames skipped after a stall and a variable frame rate. The two reviewed
works answer them with a new database and model (loose) and with a federated
scheduler for multi-view sports (distributed, \autoref{sec:distributed}).

A distinct family of supervised models targets live video streaming, where the
distortion mechanisms differ from video-on-demand. Zhu \textit{et al.}~\cite{zhu2026taolive}
report that existing \QoE{} databases, including the LIVE-NFLX and WaterlooSQoE
series, represent stalling as repeated frames under VoD setups, and therefore
cannot capture the frame skipping and variable frame rate that occur in live
broadcasts after a stall. To close this gap, the authors introduce the TaoLive
\QoE{} Database, which contains 42 source videos from the Taobao Live platform
and 1,155 distorted clips generated by combining compression, stalling,
accelerated playback, and frame skipping, with ground-truth \QoE{} scores
collected from 20 subjects. They then propose Tao-QoE, a DNN-based model that
reconstructs the video sequence from the presentation time stamps, extracts
multi-scale semantic features with a Swin Transformer, extracts motion features
from inter-frame optical flow with PWC-Net, and regresses a retrospective \QoE{}
score. The authors report that Tao-QoE attains the highest weighted-average PLCC
($0.913$) and SRCC ($0.902$) among the compared \QoE{} models across six \QoE{}
databases, and that it also outperforms four no-reference VQA models on nine UGC
VQA databases. The paper also reports a limitation: the optical-flow extraction
step consumes $77.68\%$ of the total computation time, so the full Tao-QoE model
does not run in real time, whereas the variant without the optical-flow motion
features processes $46.88$~fps.

The TaoLive study is informative for two reasons beyond the model. First, it
quantifies the perceptual saturation of resolution and frame rate: the authors
report that raising the frame rate from 20 to 25~fps produces a statistically
significant \QoE{} gain, whereas the gain from 25 to 30~fps is not statistically
significant, and that the 1080p condition scores significantly higher than the
720p condition. Second, it reports a non-trivial result on accelerated playback:
within the evaluated acceleration ratio range of 1.0 to 2.25, the decrease in
\QoE{} is not statistically significant, whereas the stalling mode itself is.
These findings support the argument in \autoref{sec:challenges-key} that
standardized benchmarks are a precondition for comparing \QoE{} models across
content types.

\emph{AR/VR and cloud gaming.} These applications face strict latency
constraints to prevent motion sickness and maintain immersion~\cite{ml-5}. High
bandwidth requirements for 360-degree video transmission challenge existing
cellular infrastructure. Most surveyed frameworks lack the ultra-low latency
processing required for effective edge-based adaptation in these scenarios. The
constraint matches the coupling found: immersive media is the only domain where
most works (6 of 11) put the model inside the loop or across nodes, because
rendering and transmission are decided per frame or per slot. All six are
evaluated in simulation or on testbeds. Cloud gaming has the same latency
constraint but no reviewed work with a model in the loop; its works are
subjective studies and offline quality
models~\cite{sr-7,rossi2025qoe,ml-1,ml-7,ml-19}. A model trained in one gaming
context fails in another~\cite{rossi2025qoe}.

A further development is orthogonal to the coupling taxonomy but material to its
evaluation: the emergence of standardized, domain-specific benchmarks. Hossain
\textit{et al.}~\cite{hossain2026vvq} release VVQ-DS, a two-phase VVC/H.266 benchmark for
volumetric video quality. Phase 1 provides 350 static and 176 dynamic sequences
with subjective baselines. Phase 2 provides 8,448 multi-view encodes generated
from 176 base sequences, four viewpoints, three compression configurations
(geometry-only, texture-only, and combined geometry-texture), and four
quantization parameters (15, 25, 35, and 45), with PSNR, SSIM, and VMAF computed
per clip. The authors report a MOS-rated subset of 240 conditions, and provide
pMOS labels for the full catalogue by learning a mapping from objective
indicators and encoding factors to MOS with an XGBoost regressor under five-fold
GroupKFold. The dataset-wide mean pMOS decreases monotonically with increasing
QP, and VMAF exhibits a stronger rank association with pMOS (SRCC $0.795$) than
PSNR ($0.617$) or SSIM ($0.589$). The authors state explicitly that the pMOS
values are model-derived and should be treated as a scalable proxy rather than a
substitute for direct human judgement. Unlike most volumetric datasets, which
follow MPEG point-cloud coding pipelines, VVQ-DS is anchored to a VVC/H.266
toolchain, and it supplies the kind of aligned objective features and
machine-readable metadata that integrated \QoE{} pipelines need for training and
for benchmarking.

\emph{Real-time interactive services.} Calls and interactive applications are
judged on delay and interactivity. Two of the three reviewed real-time
communication works put the model in the loop, as a bandwidth
estimator~\cite{wang2025llm4band} or as monitoring agents~\cite{rezaee2025realtime};
the third estimates call \QoE{} from faces offline~\cite{up-1}.

The same shift toward standardization is visible at the network plane. The ITU-T
G.1051 methodology~\cite{itut2023g1051}, evaluated by
Caso \textit{et al.}~\cite{caso2025standardized} (Section~\ref{sec:background}), provides
a standardized procedure for measuring \QoE{} of real-time interactive services
in 5G and beyond-5G systems, and it makes the reported numbers comparable across
deployments. The authors validate the methodology on the Karlstad University 5G
non-public network, and report that the 5G-enabled mode attains a higher median
i-score than LTE-only operation for the two gaming patterns, while the Industry
4.0 pattern obtains a $0\%$ i-score in every evaluated configuration. This
result is directly relevant to the integrated \QoE{} pipelines discussed in this
survey: a standardized network-plane \QoE{} KPI such as the i-score is exactly
the kind of measurement a RIC-mediated controller needs in order to verify its
decisions against an external reference, independently of the model it uses
internally.

\emph{IoT applications.} IoT domains operate under tight resource constraints
(battery, computation). Traditional \QoE{} models that require heavy client-side
processing are unsuitable for low-power IoT sensors. IoT \QoE{} is often
context-dependent rather than purely media-dependent, which requires specialized
lightweight models that few existing frameworks provide. The three reviewed IoT
works confirm this: a MOS-survey framework~\cite{sr-19}, a formula-based
resource estimator~\cite{sr-15} and one learned scheduler at the
edge~\cite{Mdemaya}.

\subsection{\underline{Model Selection}}
\label{sec:synthesis-decision}\label{sec:ml-vs-stat}

ML models outperform statistical methods in dynamic network environments because
they capture non-linear QoS-to-\QoE{} relationships that linear regression and
time-series models cannot represent~\cite{ml-9,ml-6}. An improved TCN infers in
under 1~ms per sample on a PC and on a phone~\cite{ml-18} and RFR models map
radio metrics (RSSI, RSRP, RSRQ, CINR) to MOS, PSNR and frame
delay~\cite{d1}. However, ML requires large labeled datasets, has high training
cost, and reduces interpretability. Statistical methods remain preferable where
data is scarce or explainability is required. The practical choice depends on the
latency budget, the dataset size, and the operator's interpretability
requirements.

The following guide is built only from results verified in the reviewed works.
It is a starting point for model selection, not a ranking, because the works
report different metrics on different datasets.

\begin{itemize}
    \item \emph{Latency-critical (milliseconds-scale):} use an improved TCN,
          which infers in under 1~ms and three times faster than an
          LSTM~\cite{ml-18}. Among tree ensembles, XGBoost predicts about 25
          times faster than RF in the framework of~\cite{mine-archive}, at a
          lower $R^2$ (0.78 against 0.968). RFR suits offline estimation from
          network KPIs~\cite{ml-3}. Avoid deep recurrent networks (LSTM) because
          of inference latency.
    \item \emph{Small or sparse datasets:} use SVR~\cite{ml-1} or
          RBFN~\cite{ml-10}. Deep learning models (DNN~\cite{ml-9}) risk
          overfitting without large training sets; on about 22,000 samples, RF
          reaches $R^2 = 0.968$ against 0.75 for a DNN~\cite{mine-archive}.
    \item \emph{Complex non-linear relationships:} use DNN~\cite{ml-9} or a
          recurrent model (LSTM~\cite{ml-6}, BLSTM~\cite{ml-21}). Statistical
          methods such as linear regression cannot represent these dynamics.
    \item \emph{Unlabeled data:} use unsupervised learning (content
          clustering~\cite{ml-20}) for clustering and anomaly detection.
    \item \emph{Explanation required:} use an inherently interpretable model
          such as TSKAN~\cite{tskan2026}, or SHAP and Conformal Prediction over
          tree ensembles~\cite{bilbao2025benchmarking,osee2025hybrid}
          (\autoref{sec:challenges-key}).
    \item \emph{Live control in the network:} use an RL controller in the tight
          or distributed pattern (\autoref{sec:tight},
          \autoref{sec:distributed}); all such controllers reviewed here are
          evaluated in simulation or on testbeds, except the CFA
          pilot~\cite{sr-3}.
\end{itemize}

\subsection{\underline{Summary and Lessons Learned}}
\label{sec:domains-summary}

\autoref{tab:domains} gives the answer to \texttt{RQ3} in counts. Three lessons follow.
First, the coupling follows the decision deadline of the service: per-frame
immersive delivery is where models move into the loop, while streaming, with
seconds of buffer, keeps most models offline. Second, cloud gaming and IoT are
under-served: both have strict constraints and almost no learned controller.
Third, results cannot be compared across domains, because each domain uses its
own datasets and metrics; the benchmark problem that this raises is taken up in
\autoref{sec:challenge}.

\section{Open Challenges and Emerging Directions}
\label{sec:challenge}

\begin{figure}[]
    \centering
    \includegraphics[width=0.9\linewidth, trim={0.5cm 0.5cm 0.5cm 0.5cm}, clip]{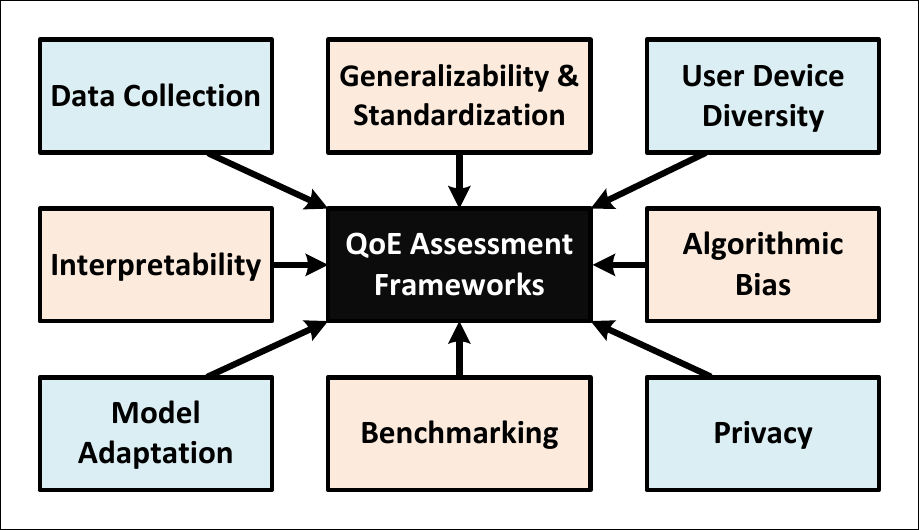}
    \vspace{-1em}
    \caption{Key challenges in the design and development of \QoE{} assessment
    frameworks.}
    \label{fig:challenges}
\vspace{-1em}
\end{figure}

This section answers \texttt{RQ4}: \emph{which problems remain open?} It covers stage~6 of
\autoref{fig:all-in-one}, the feedback that should keep a deployed model
accurate. This stage is the least developed: only LingXi adapts its objective
from live feedback in production~\cite{jia2025lingxi}. Drift detection with
self-healing is tested only in trace replay~\cite{selma2026qoe}.
\autoref{sec:synthesis-generations} places the reviewed work in three
generations. \autoref{sec:challenges-key} reviews the challenges that cut across
domains and call for shared methods, such as data collection, benchmarking and
privacy; the constraints that call for domain-specific design were treated in
\autoref{sec:domains}. \autoref{sec:emerging} names the directions that rest on
a single work so far. \autoref{fig:challenges} summarizes the cross-domain
challenges.

\subsection{\underline{Generational Evolution of Technical Solutions}}
\label{sec:synthesis-generations}

The evolution of \QoE{} solutions follows three generations.
\emph{Gen 1 (2010--2015): static measurement} focuses on statistical analysis,
subjective surveys and the first offline models, such as the neural network
of~\cite{sr-11}~\cite{sr-8}. \emph{Gen 2 (2016--2020): reactive ML} adopts
supervised learning (DNN, CNN) to predict \QoE{} after
transmission~\cite{ml-9,ml-6}. \emph{Gen 3 (2021--2026): proactive and
generative} shifts toward TCNs, GANs, continual learning, and distributed
integration to forecast and enhance \QoE{} before issues
occur~\cite{ml-18,ml-19,up-2,zhou2025federated}. Recent work integrates LLMs for
subjective analysis and bandwidth estimation~\cite{panahi2025bridging,wang2025llm4band},
which brings language models into both \QoE{} assessment and network control.
Emerging 6G architectures use SDN/NFV, MEC, and AI/ML for adaptive video
streaming with enhanced \QoE{} optimization~\cite{alsader2025qoe}.

Across the surveyed ML studies, results are reported with different metrics
(accuracy, PCC, LCC, $R^2$, RMSE) on different datasets, so they cannot be
ranked against each other; Tables~\ref{tab:summl} and~\ref{tab:summl-rl} give
each result with its metric. Performance is highly task- and
dataset-dependent. Of the 68 reviewed works, video streaming accounts for 31,
immersive media (AR/VR, 360-degree and volumetric video) for 11, gaming video and
cloud gaming for 5 and IoT for 3 (\autoref{tab:domains}). AR/VR and IoT drive the
shift toward Gen 3 solutions because of their strict latency and resource
constraints.

\subsection{Key Challenges}
\label{sec:challenges-key}

\subsubsection{Data collection}
Large-scale collection faces restricted access, user
unwillingness, and high costs~\cite{sr-18,sr-6,f4-8}. Open-source tools that
rely on voluntary participation collect less diverse data than closed-source
solutions with dedicated infrastructure~\cite{sr-10,sr-9}. Data quality issues
and non-stationary environments such as moving users and changing cells further
complicate training.

\subsubsection{Generalizability}
Methods are application-specific and lack unified
standards~\cite{sr-10,sr-14,sr-16,sr-18}. Most \QoE{} data collection is scoped
to single services such as YouTube, and the absence of common metric definitions
makes results across studies incomparable. Cross-service and cross-network
standardization is needed~\cite{sr-6,sr-9}.

A production measurement now anchors this challenge on the optimistic side. Jia
\textit{et al.}~\cite{jia2025lingxi} report that system-wide linear \QoE{} objectives are
nearing their performance ceiling, with gains of only $0.5\%$ to $2\%$ between
algorithm variants in production tests, while a user-level objective,
personalized through an exit-rate predictor and tuned online, delivers a further
$+0.146\%$ in total watch time and a $-1.287\%$ decrease in stall time. The same
study reports substantial heterogeneity in stall tolerance: approximately $20\%$
of users exhibit minimal tolerance to stall events, another $20\%$ tolerate
stalls longer than $5$~s, and about $10\%$ tolerate stalls exceeding $10$~s. The
generalizability problem is therefore not that user-level models fail to
generalize; it is that a model generalized across users underperforms one fitted
to the individual user, and the individual-level signal is available only in a
deployed system.

\subsubsection{Device diversity}
The heterogeneity of hardware and software across user
devices increases \QoE{} modeling complexity and makes cross-device prediction
and comparison challenging~\cite{sr-21,sr-6}. Device-aware models and further
research into heterogeneity-induced quality issues are required.

\subsubsection{Interpretability}
Deep learning models are opaque and cannot explain their
\QoE{} predictions to network
operators~\cite{kougioumtzidis2022survey,nasralla2023exploring}. The
accuracy--interpretability trade-off is particularly acute for CNNs and LSTMs.
However, recent research has begun to bridge this gap. Wehner et
al.~\cite{wehner2026tutorial} show how XAI methods reveal the influence factors
behind data-driven \QoE{} models and build trust in their predictions. Three
recent works, described next, address it directly.

\subsubsection{Explainable and interpretable \QoE{} models}
The black-box nature of
deep learning models (\textit{e.g.,} LSTMs and Transformers) limits their adoption in
operational networks where operators require explanations for \QoE{}
predictions. Wehner \textit{et al.}~\cite{wehner2026tutorial} formalize this need in a
tutorial on data-driven \QoE{} modeling with XAI, which applies post-hoc
explainers and interpretable models to video streaming and web \QoE{} data. To
address this, recent works have focused on inherently interpretable models and
on explaining black-box models after training. TSKAN~\cite{tskan2026}
introduces an interpretable architecture that combines Kolmogorov-Arnold
Networks (KANs) with frequency-domain features (extracted via DFT). By learning
transparent univariate functions, TSKAN achieves an RMSE of 0.2169 on the
LIVE-Netflix-II dataset (on z-scored MOS), outperforming other interpretable
baselines (\textit{e.g.,} explainable boosting machine and neural additive models) while
maintaining transparency. Furthermore, Bilbao et
al.~\cite{bilbao2025benchmarking} systematically benchmark ML models for video
streaming \QoE{}, evaluating not only predictive accuracy (MAE and SMAPE) but
also computational efficiency (training/inference time) and prediction confidence
(via Conformal Prediction). They demonstrate that ensemble-based randomization
networks (edRVFL) can achieve high accuracy with low training times, while SHAP
analysis confirms that features like TCP connection time and ICMP round-trip
time strongly influence the predicted buffering ratio when the training set is
small. Finally, Osee \textit{et al.}~\cite{osee2025hybrid} propose a hybrid
ensemble framework (XGBoost and Random Forest with a meta-learner) that
integrates QoS metrics with subjective user satisfaction surveys. Using SHAP
analysis, they identify user-rated factors such as image quality and message
delivery speed as the strongest predictors. They argue that hybrid models
combining network KPIs with user feedback provide actionable insights for
operators, particularly in resource-constrained environments. Despite these
advances, feature-importance methods and explainability techniques are still
needed for large-scale, real-time deployments.

\subsubsection{Algorithmic bias}
Training data collected from non-representative
demographic or network conditions propagates bias into \QoE{} models, which
affects fairness~\cite{kougioumtzidis2022survey,sr-17,sr-19}. Standardized test
sets, bias quantification metrics, and mitigation techniques remain open
research problems.

\subsubsection{Model adaptation}
Models trained on historical data become outdated as
network conditions, display technologies, codec standards, and user expectations
evolve~\cite{kougioumtzidis2022survey,ml-9}. Continual and incremental learning
approaches that update models from live streaming sessions are
required~\cite{nasralla2023exploring,sr-18}. This challenge becomes even more
acute for emerging interactive AI systems. P{\'e}rez \textit{et al.}~\cite{perez2026video}
highlight that moving from video \QoE{} to LLM \QoE{} requires a fundamental
reformulation: the stimulus becomes a multi-turn interaction rather than static
content, and technical constraints shift from network metrics toward compute and
memory budgets. Furthermore, Karapantelakis \textit{et al.}~\cite{karapantelakis2026qoe}
note that in 5G networks, end-to-end encryption limits observability, which
reduces the accuracy of traditional \QoE{} assessment methods. To address this,
they propose a multi-agent system where an LLM-driven Deployment Agent translates
high-level user-level \QoE{} objectives into measurable network and cloud
performance rules, which are then enforced by a Monitoring Agent that triggers
corrective actions. These works suggest that future \QoE{} management has to not
only adapt to network changes but also to the dynamic and opaque nature of
AI-driven applications.

Topology change is a second form of model drift. Chen et
al.~\cite{chen2026continual} report that when the number of ground users
increases and the UAV-GU topology changes, conventional FoV-aware DRL agents
suffer catastrophic forgetting and converge to sub-optimal policies; their
DGC-DRL framework mitigates this with a dynamic graph representation and a
continual-learning module, but the solution assumes the controller can be
re-trained at each topology change, which is itself a deployment constraint.

\subsubsection{Drift-aware and self-healing \QoE{} prediction}
A complementary line
addresses the degradation of \QoE{} predictors after deployment. Selma et
al.~\cite{selma2026qoe} propose QoE-Foresight, a closed-loop framework that
combines a multimodal feature pipeline (EMMA), a regularized Ridge \QoE{}
predictor, a Hybrid Drift Detection Engine (HDDE) combining HDDM-A with an
uncertainty-aware drift filter (UADF) and a Kolmogorov--Smirnov residual test, a
double deep Q-Network (DDQN) self-healing controller, and a SHAP-based
explainability module. On an 813-sample fused benchmark drawn from four public
sources and synthetic device traces, the Ridge predictor reports
$R^2 \approx 0.953$ with $0.12$~ms inference latency, the HDDE reports an
F1-score of $0.967$ with an average detection delay of $0.035$~s, and the
self-healing controller reports a mean \QoE{} gain of $0.73$ MOS under controlled
drift-replay scenarios. The authors state explicitly that the self-healing
experiments use offline trace replay with controlled drift injection rather than
a live production system, so the recovery figures are a proof-of-concept rather
than a deployment measurement.

Drift-aware self-healing systems respond to this challenge by monitoring the
\QoE{} predictor's residuals online and triggering a corrective action when the
residual distribution departs from its nominal regime.
QoE-Foresight~\cite{selma2026qoe} reports its gains under controlled
drift-replay, and the same work notes that its self-healing controller is trained
on an abstracted action space and has not been coupled to a live ABR engine or a
production CDN. The challenge of adapting deployed \QoE{} models therefore
remains open at the level of end-to-end deployment, not only at the level of
drift detection.

\subsubsection{Benchmarking}
Isolated contributions evaluated on custom datasets prevent
fair comparison~\cite{kougioumtzidis2022survey}. Standardized expert-labeled
datasets across diverse contexts, particularly for gaming and AR/VR, are needed
to assess real progress in ML-based \QoE{} modeling. Three developments narrow
this challenge without eliminating it; their details are given in
\autoref{sec:domains}. At the media plane, Zhu \textit{et al.}~\cite{zhu2026taolive}
introduce TaoLive, which they describe as the first live-streaming \QoE{}
database, and show that existing \QoE{} databases under-represent the frame
skipping and variable frame rate of live broadcasts. At the immersive plane,
Hossain \textit{et al.}~\cite{hossain2026vvq} release a factorial VVC/H.266 volumetric
benchmark with 8,448 multi-view encodes and aligned objective metrics, but note
that their catalogue-wide pMOS labels are model-derived from only 240
human-rated conditions. At the network plane, the ITU-T G.1051
methodology~\cite{itut2023g1051}, as evaluated by
Caso \textit{et al.}~\cite{caso2025standardized}, provides a standardized procedure for
measuring the \QoE{} of real-time interactive services, but its interactivity
score is defined on traffic patterns rather than on the rendered output a user
actually sees. No single current benchmark covers measurement, content, and
model evaluation under one protocol, which is the residual gap for benchmarking
in integrated \QoE{} systems.

Two further results show why this gap matters. Rossi \textit{et al.}~\cite{rossi2025qoe}
evaluate a \QoE{} model trained on mobile cloud gaming against a new PC cloud
gaming dataset collected from 30 participants over 15 network conditions, and
find significant prediction errors, which shows that a model trained in one
context does not transfer directly to another. Wu \textit{et al.}~\cite{wu2025qoe} report
that public datasets with synchronized visual and haptic feedback are scarce, so
they build an HDT multi-modal interaction platform and release the resulting
dataset themselves. Both cases indicate that the benchmark gap is not only about
the absence of standardized protocols; it is also about the absence of datasets
that cover the modality combinations and delivery contexts the models are now
expected to handle.

\subsubsection{Privacy}
The general data protection regulation (GDPR) and the California
consumer privacy act (CCPA) restrict subjective \QoE{} data collection, and
enhanced encryption hinders real-time packet inspection. Federated learning,
on-device processing, and anonymization are needed to balance \QoE{} model
accuracy with regulatory compliance. However, even federated learning faces
trade-offs between privacy and model
utility~\cite{garcia2025decentralized}. Garcia et
al.~\cite{garcia2025decentralized} propose combining Fully Homomorphic Encryption
(FHE) with blockchain-based smart contracts for 6G networks, allowing
computations on encrypted \QoE{} data without exposing raw user information or
proprietary business data.

This survey also notes a possible architectural route to privacy at the RAN. In
the O-RAN-integrated playback framework of Wen \textit{et al.}~\cite{wen2026immersive},
the per-frame controller runs as a Near-RT RIC xApp on operator infrastructure,
so control decisions are taken inside the operator domain. Wen \textit{et al.} do not
study privacy, however. Their framework still forwards multi-view uploads to the
content provider. The privacy question for integrated \QoE{} systems therefore
has one tested answer, cryptographic protection (FHE and blockchain), and one
untested option, operator-side control.

\subsection{\underline{Emerging Directions}}
\label{sec:emerging}

Five directions rest on a single reviewed work each, so this survey records them
as emerging directions rather than as categories. Three refine the existing
patterns: \emph{objective tuning} under loose coupling, where LingXi tunes the
objective of a production ABR algorithm (\autoref{sec:loose}); \emph{RIC-mediated
control}, where a RIC xApp decides radio, power and compute per frame
(\autoref{sec:distributed}); and \emph{federated scheduling}, where edge nodes
learn a shared stream scheduler (\autoref{sec:distributed}). Two fall outside
the patterns, which were formulated for video-only delivery on fixed
infrastructure: multi-modal integration and mobile-edge integration, described
next.

Two further developments are emerging directions that the current taxonomy does
not yet cover. The first is multi-modal \QoE{}. Wu \textit{et al.}~\cite{wu2025qoe} show
that when the \QoE{} model integrates visual and haptic feedback, the controller
has to jointly set the bandwidth split, the video bitrate and rendering mode, and
the haptic compression threshold under a synchronization constraint between the
two streams. The coupling between the two modalities is not represented in the
loose, tight, or distributed/federated patterns, which were formulated for
video-only pipelines. The HDT case is therefore named here as \emph{multi-modal
integration}, with the VisHap controller as its current example.

The second is the UAV edge. Chen \textit{et al.}~\cite{chen2026continual} consider UAVs
that capture, encode and transmit the content. A ground controller collects the
network state and trains the policy, while each UAV runs the trained policy for
scheduling and power allocation, so the inference moves with the platform, and
the set of users changes over time. The distributed/federated pattern assumes a
fixed infrastructure; the UAV case does not. The authors handle the churn with a
dynamic heterogeneous graph and a continual-learning module, and the resulting
architecture is an instance of what this survey terms \emph{mobile-edge
integration}. This survey treats both as emerging directions rather than as
separate patterns, because each currently rests on a single study. They mark
where the taxonomy may need to grow as the field moves beyond video-only
delivery on fixed infrastructure.

A further direction changes the service rather than the coupling: the \QoE{} of
interactive LLM services, for which the position paper of~\cite{perez2026video}
sets out the gaps (\autoref{sec:challenges-key}).

In parallel, Lyu \textit{et al.}~\cite{lyu2026drllms} address the \QoE{} of interactive
LLM streaming services themselves. They introduce DRLLMS, which treats
chain-of-thought (CoT) length as a control knob to optimize \QoE{}. By
dynamically adjusting CoT length based on network conditions (using a
lightweight grid-search optimizer), they balance responsiveness (measured by
Time-to-First-Answer, TTFA) and answer quality, improving user experience across
diverse bandwidth scenarios.

\subsection{\underline{Summary and Lessons Learned}}
\label{sec:challenge-summary}

Three lessons answer \texttt{RQ4}. First, the feedback stage of \autoref{fig:all-in-one}
is the least developed: drift handling and self-healing are tested only in trace
replay. Only LingXi adapts from live feedback in production. Second, the
benchmark gap is now a gap of coverage: new databases exist for live streaming,
volumetric video and network interactivity, but none covers measurement, content,
and model evaluation under one protocol. Third, four of the five emerging
directions rest on simulation or testbeds alone. Each needs a second study before
it can become a category. The conclusion that follows answers the four questions
in order.

\section{Conclusion}
\label{sec:conc}

This survey asked four questions about how \QoE{} measurement frameworks and ML
models work together, and answered them on a documented set of 68 works. For
\texttt{RQ1}, the coupling is mostly loose: most ML-based works train and test
offline on stored data, and only a minority run a model inside the live loop or
distribute training across nodes. Most measurement frameworks train no model
themselves. For \texttt{RQ2}, a smaller set of works act on the service through
three enhancement mechanisms, of which only predictive optimization of the
bitrate or ABR configuration has production evidence, from CFA and LingXi.

For \texttt{RQ3}, the coupling follows the decision deadline of the service.
Immersive media, with per-frame decisions, holds most of the tight and
distributed controllers; video streaming, buffered over seconds, keeps most
models offline; cloud gaming and IoT have strict constraints but few learned
controllers. Reported results use different metrics, so the model-selection
options are ordered by latency, dataset size and complexity rather than ranked.
For \texttt{RQ4}, feedback is the weakest stage of the loop: drift handling is
tested in trace replay only. Several challenges remain open, and five emerging
directions rest on a single work each: objective tuning, RIC-mediated control,
federated scheduling, multi-modal integration and mobile-edge integration.

Three generational phases are traced: static measurement, reactive ML, and
proactive and generative solutions, the last including LLM integration. Two
directions now shape the frontier of Gen~3: explainable \QoE{} models and
federated or privacy-preserving \QoE{} systems. Future work should move
drift-aware self-healing from trace replay to live deployments, and extend
multi-modal and mobile-edge integration beyond single studies.

\bibliographystyle{IEEEtran}
\bibliography{References.bib}

\vspace{-4em}
\begin{IEEEbiography}[{\includegraphics[width=1in,height=1.25in,clip,keepaspectratio]{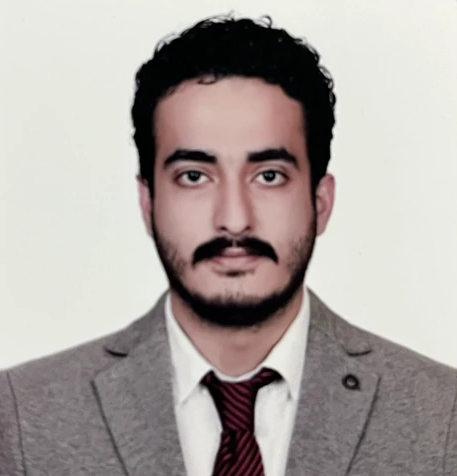}}]{Parsa Hassani Shariat Panahi}
received his B.Sc. degree in Computer Engineering from Azad University South Tehran Branch, Tehran, Iran, and the M.Sc. degree in Computer Engineering - Computer Networks from Iran University of Science and Technology, Tehran, Iran. His research interests include cellular networks, QoE assessment, telecommunication networks, wireless communication, and machine learning. He can be reached at parsa\_hassani@comp.iust.ac.ir.
\end{IEEEbiography}
\vspace{-4em} 
\begin{IEEEbiography}[{\includegraphics[width=1in,height=1.25in,clip,keepaspectratio]{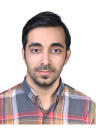}}]{Amir Hossein Jalilvand}
received his B.Sc. degree in Computer Engineering from Bu-Ali Sina University, Hamadan, Iran, and the M.Sc. degree in Computer Engineering - Computer Architecture from Iran University of Science and Technology, Tehran, Iran. He is currently pursuing his Ph.D. in Computer Engineering. His research interests include cellular networks, stochastic and unary computing, computer architecture, fuzzy logic, and machine learning. Mr. Jalilvand has authored several publications in these fields. He can be reached at jalilvand\_a@comp.iust.ac.ir.
\end{IEEEbiography}
\vspace{-4em} 
\begin{IEEEbiography}[{\includegraphics[width=1in,height=1.25in,clip,keepaspectratio]{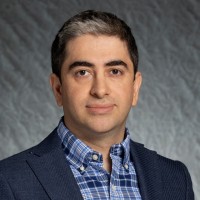}}]{M. Hassan Najafi}
received his Ph.D. in electrical and electronics engineering from the University of Minnesota-Twin Cities, Minneapolis, MN, USA, in 2018. He is currently an Associate Professor at the Electrical, Computer, and Systems Engineering Department at Case Western Reserve University. His research interests include stochastic and approximate computing, unary processing, in-memory computing, and hyperdimensional computing. He has authored/coauthored more than 110 peer-reviewed papers and has been granted 10 U.S. patents with more pending. Dr. Najafi received the NSF CAREER Award in 2024, the Best Paper Award at GLSVLSI'23 and ICCD'17, and the 2018 EDAA Outstanding Dissertation Award. Dr. Najafi is a senior member of IEEE and a senior member of NAI. He can be reached at najafi@case.edu.
\end{IEEEbiography}

\end{document}